\documentclass[a4paper,11pt]{article}
\usepackage{jheppub} 
\usepackage{lineno}
\usepackage{comment}
\usepackage{placeins}
\usepackage[all]{nowidow} 

\usepackage{parskip}

\usepackage{siunitx}

\usepackage{graphicx}
\usepackage{subcaption}
\usepackage{pdfpages}
\usepackage{multirow}
\usepackage{xspace}
\usepackage[normalem]{ulem}
\newcommand{\pT}{\ensuremath{p_{\text{T}}}\xspace}

\arxivnumber{1234.56789} 

\title{Enabling searches for long-lived particles at a future 10 TeV Muon Collider}

\author{M. Littmann,}
\author{M. Larson,}
\author{K. Wandrisco,}
\author{B. Rosser,}
\author{T. Flicker,}
\author{K. Huang,}
\author{L. Rozanov,}
\author{and K. F. Di Petrillo}
\affiliation{The Enrico Fermi Institute, The University of Chicago,\\
Chicago, IL, USA}

\emailAdd{karri@uchicago.edu, brosser@uchicago.edu}

\abstract{Muon colliders offer fantastic opportunities to explore new phenomena at the energy frontier. However, beam-induced-backgrounds from muon decays pose significant challenges for detector design, readout, and reconstruction. 
Previous detector studies have employed stringent hit-timing requirements to reduce occupancy to manageable levels with negligible efficiency loss for prompt Standard Model particles. In the spirit of maximizing discovery potential, we investigate the capability of detecting meta-stable charged long-lived particles at a 10 TeV muon collider. As a benchmark, we consider a Gauge Mediated Supersymmetry Breaking (GMSB) model in which the supersymmetric tau partner is long-lived and can be identified as a high momentum, slowly moving track. We find that nominal hit-timing selections are too restrictive, and investigate the impact of looser requirements. We demonstrate that it is possible to recover sensitivity to particles with masses close to $\sqrt{s}/2$ by loosening hit-timing criteria. We find track-quality, timing, and mass-dependent kinematic selections strongly suppress reconstructed BIB and muon backgrounds, while the ditrack invariant mass provides additional separation between signal and BIB. These results motivate projected sensitivity estimates under a target zero-background reconstruction scenario.}

\begin{document}
\maketitle
\flushbottom

\section{Introduction}
\label{sec:intro}
A 10 TeV scale muon collider offers an attractive option for the future of the energy frontier. Muons combine the clean environment characteristic of lepton colliders with center-of-mass energies comparable to those of hadron colliders in a compact and power-efficient machine. Such a facility is particularly well-suited to probe dark matter, naturalness, and electroweak symmetry breaking~\cite{towards-mucoll,smashers-guide}. 

However, muon decays near the interaction point pose several challenges for detector design, reconstruction, and physics performance. Current detector designs include tungsten nozzles placed in the forward region to prevent high energy muon decay products from entering the detector \cite{Ally:2022rgk, maia-concept, music, imcc-euro-strat}. The resulting showers lead to $\mathcal{O}(10^8)$ low energy neutrons, photons, and electrons leaking into the detector each bunch crossing. These beam-induced-background (BIB) particles generally do not point back to the interaction region and tend to produce hits which are out of time with respect to particles produced in collisions. 

Activity due to BIB is highest in regions of the detector closest to the beamline, calling for a highly granular silicon tracker capable of precision timing measurements in every layer. Rejecting out of time hits from BIB is essential to reduce occupancy to manageable levels for reconstruction. Timing selection criteria have been previously optimized for promptly produced Standard Model (SM) particles, but will pose challenges for long-lived particle signatures such as slowly moving or displaced tracks. 

In the spirit of maximizing discovery potential, we investigate the capability to detect long-lived particles at a 10 TeV muon collider. Previous muon collider studies include a proof-of-concept search for disappearing tracks~\cite{higgsino}. Here, we focus on meta-stable charged particles that traverse the entire tracker and can be reconstructed as high momentum tracks with $\beta=v/c<1$. In comparison to other long-lived signatures, this signature produces hits with the maximum possible delay in each layer of the tracker. We quantify the impact of loosened timing criteria on detector occupancy, track reconstruction, and sensitivity to new physics. We also identify potential strategies for further improvements.

\section{Simulated datasets}
\label{sec:datasets} 

\begin{figure}[h!]
    \centering
    \includegraphics[width=0.5\textwidth]{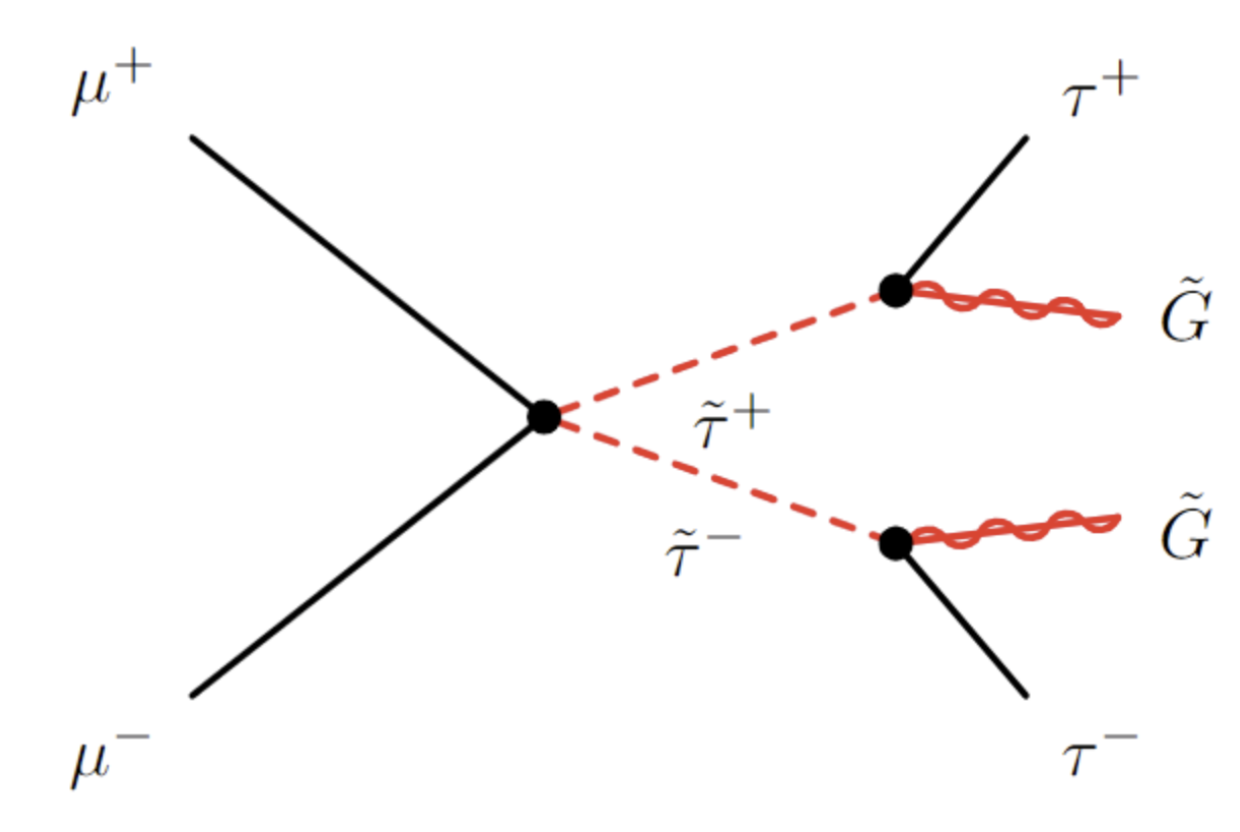}
    \caption{Benchmark model. Pair-production of long-lived staus in a simplified GMSB SUSY model, with each stau decaying to a tau and gravitino.}
    \label{fig:feynman}
\end{figure}

We consider a simplified gauge-mediated supersymmetry (SUSY) breaking model as a benchmark~\cite{Giudice:1998bp,Ambrosanio:2000ik}. In this model, a nearly massless gravitino ($\tilde{G}$) is the lightest SUSY particle, while the stau ($\tilde{\tau}$) is the next-to-lightest SUSY particle. We consider exclusively $s$-channel production of stau pairs, followed by the decay $\tilde{\tau}\rightarrow\tau\tilde{G}$, with a lifetime that depends on the scale of SUSY breaking, as shown in Figure~\ref{fig:feynman}.

\begin{figure}[h!]
    \centering
    \includegraphics[width=0.65\textwidth]{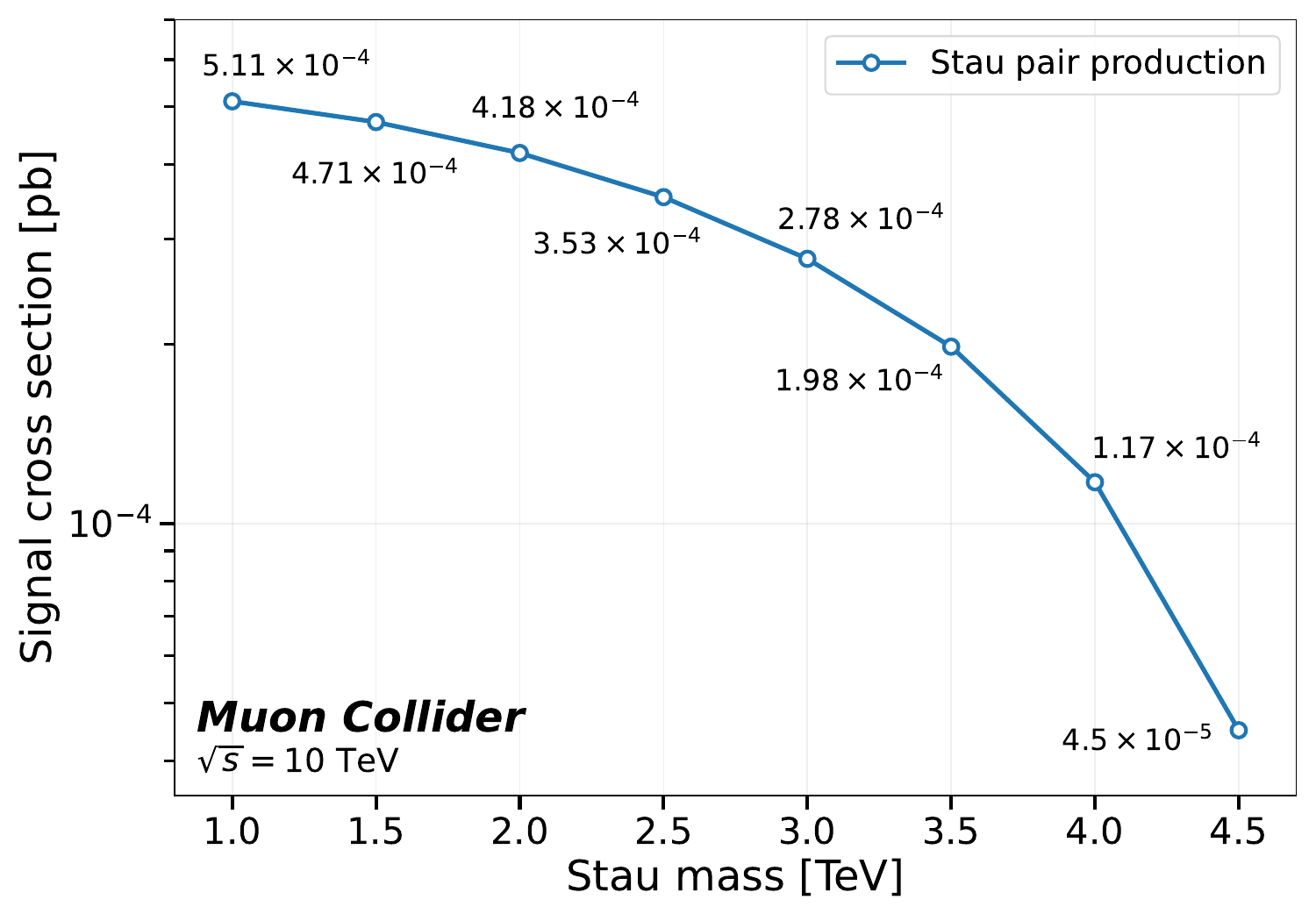}
    \caption{Leading order cross sections for stau pair-production calculated with \textsc{MadGraph5\_aMC@NLO} as a function of stau mass at \(\sqrt{s}=10~\mathrm{TeV}\) at a muon collider.}
    \label{fig:cross-sections}
\end{figure}

\begin{figure}
    \centering
    \includegraphics[width=0.48\linewidth]{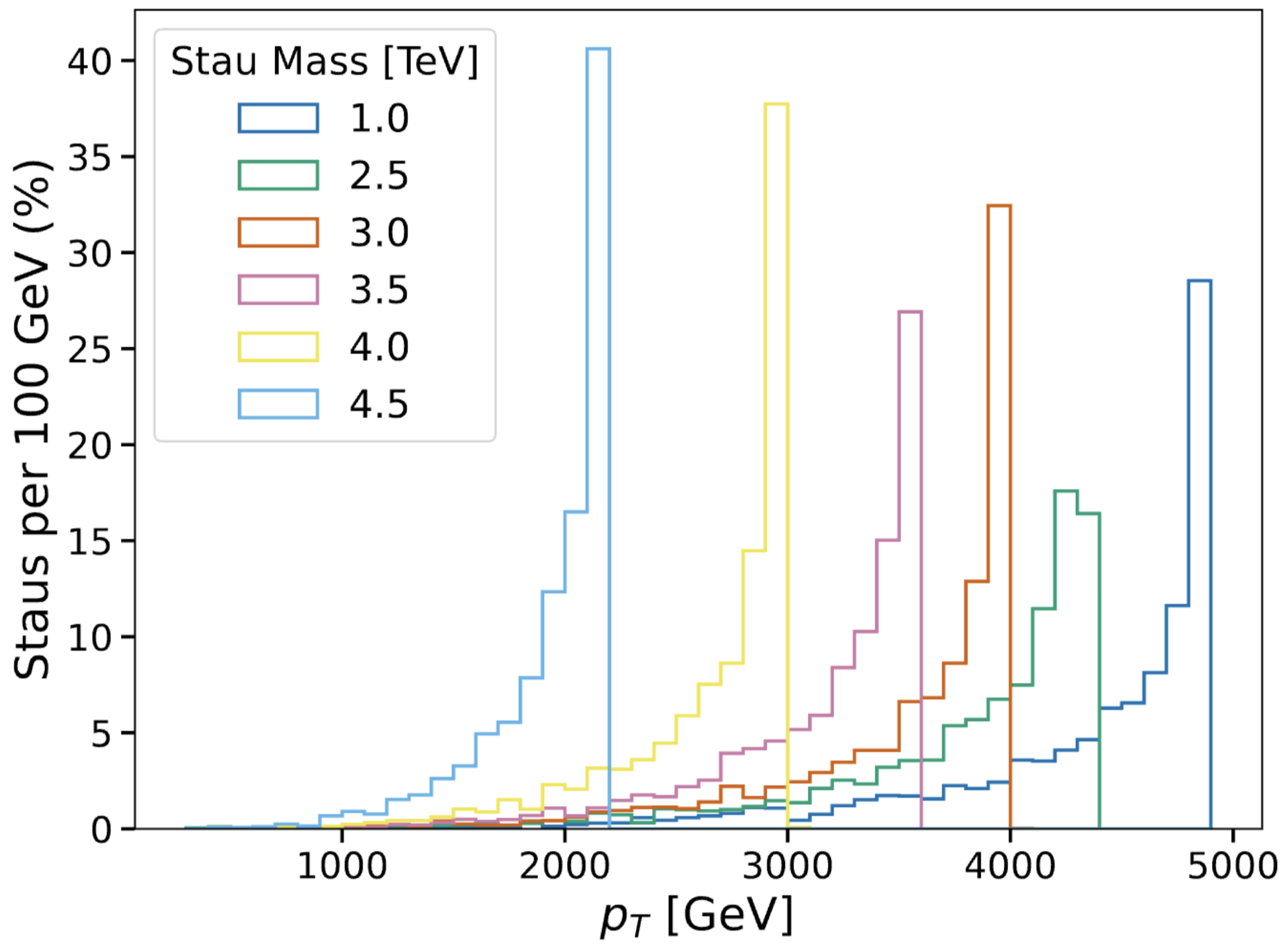}
    \includegraphics[width=0.48\linewidth]{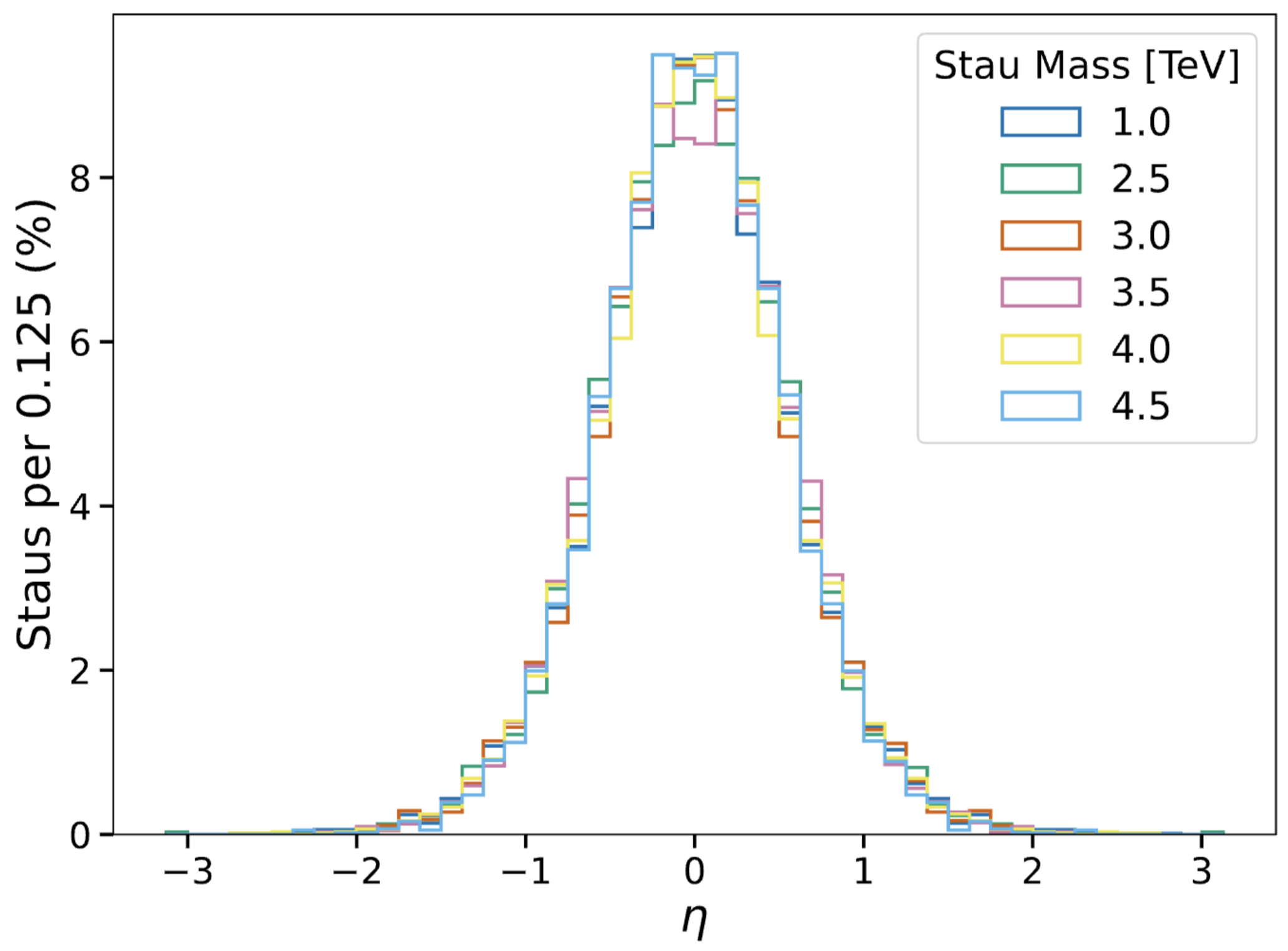}
    \caption{Kinematic properties of simulated staus including \pT{} (left), $\eta$ (right).}
    \label{fig:sim-stats}
\end{figure}

Signal events are generated at $\sqrt{s}=10~\mathrm{TeV}$ in \textsc{MadGraph5\_aMC@NLO} interfaced with \textsc{Pythia8}~\cite{madgraph,Sjostrand:2007gs}. The stau mass is varied from $1$~to~$4.5~\mathrm{TeV}$ in steps of $500~\mathrm{GeV}$. The stau lifetime is varied from 0.1 ns to 30 ns, with an additional stable benchmark. Expected cross sections are shown in Figure~\ref{fig:cross-sections}, and characteristic kinematic properties are shown at truth-level in Figure~\ref{fig:sim-stats}.


Events are simulated with BIB using software maintained by the International Muon Collider Collaboration (IMCC) \cite{bib}. BIB interactions up to the entrance of the detector region are simulated with \textsc{FLUKA}~\cite{Ahdida:2022gjl,Ballarini:2024isa}, assuming a collider lattice designed for $\sqrt{s}=10$ TeV collisions, tagged v04 by the IMCC~\cite{Skoufaris:2022wwg}. These samples are simulated independently of signal and provided by the IMCC to the larger community. 

\begin{figure}[h!]
    \centering
    \includegraphics[width=0.8\textwidth]{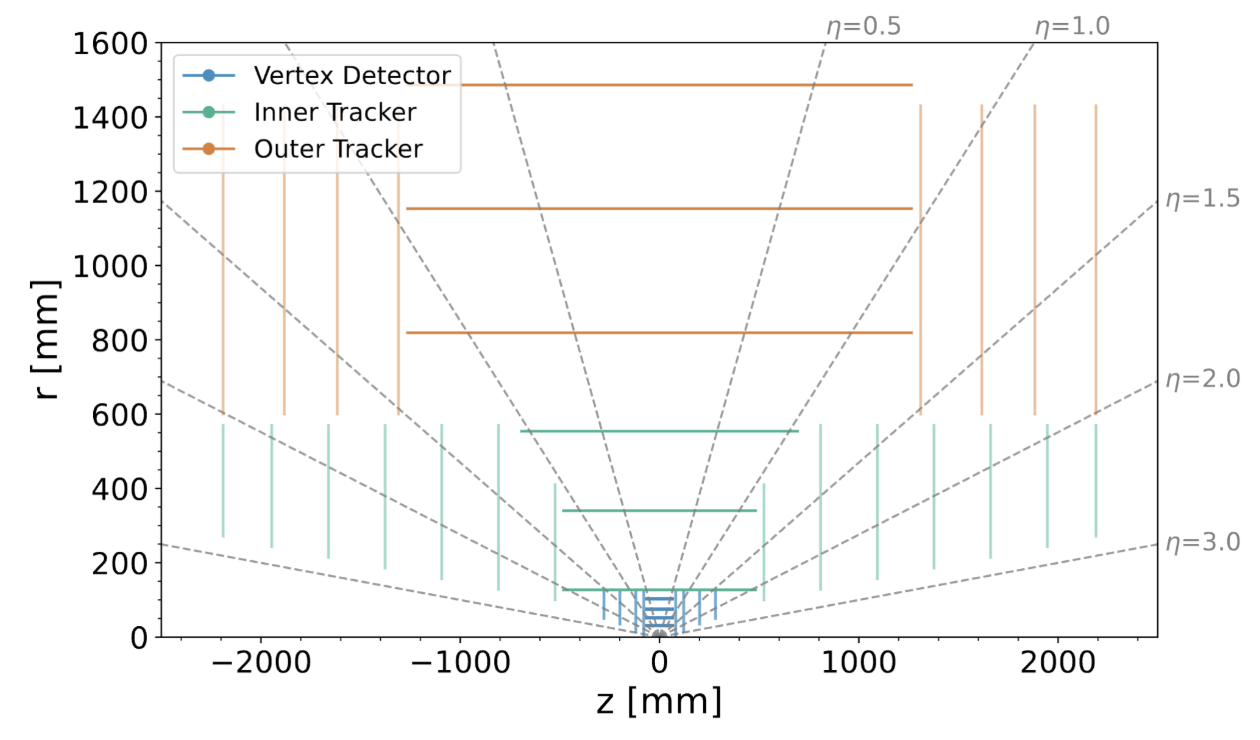}
    \caption{The Tracker geometry from \textsc{MuColl\_v1}, indicating the position of each layer in the $r{-}z$ plane~\cite{towards-mucoll}. }
    \label{fig:tracker-geo}
\end{figure}

The \textsc{MuColl\_v1} detector geometry is implemented in simulation using the DD4hep framework~\cite{dd4hep} following the design described in Ref.~\cite{towards-mucoll}. Originally designed for collisions at $\sqrt{s}=3~\mathrm{TeV}$, the Tracker and Calorimeter are surrounded by a solenoid that produces a $3.57~\mathrm{T}$ magnetic field. The Tracker, pictured in Figure~\ref{fig:tracker-geo}, consists of Vertex, Inner, and Outer sub-detectors. The Vertex Detector consists of 4 double-sided barrel layers and 4+4 endcap disks of $\SI{25}{\micro\meter}\times\SI{25}{\micro\meter}$ silicon pixels, with a time resolution of $\SI{30}{\pico\second}$. The Inner Tracker has 3 barrel layers and 7+7 endcap disks, made of $\SI{50}{\micro\meter}\times\SI{1}{\milli\meter}$ macro-pixels. The Outer Tracker has 3 barrel layers and 4+4 endcap disks of $\SI{50}{\micro\meter}\times\SI{1}{\milli\meter}$ macro-pixels. Both the Inner and Outer Tracker have a time resolution of $\SI{60}{\pico\second}$. 

The passage of signal and BIB particles through the detector are simulated separately with \textsc{GEANT4}~\cite{GEANT4:2002zbu}. Simulated signal hits are then overlaid with BIB and digitized using a key4hep~\cite{k4MarlinWrapper} based software framework. Gaussian smearing is applied to simulated hits to account for detector temporal and spatial resolution. Digitized hits are required to have a time of arrival consistent with particles originating from the interaction point in order to be considered for track reconstruction. For this study, the timing windows used in digitization are loosened to investigate sensitivity to long-lived particles.

Track reconstruction is performed with A Common Tracking Software (ACTS) version 32.1.0~\cite{ACTS}. Track seeds are formed using triplets of hits in the Vertex Detector. A minimum transverse momentum requirement of $p_\mathrm{T}>10~\mathrm{GeV}$ is applied to track seeds in order to minimize random combinations from BIB hits. ACTS propagates track seeds to find hits consistent with the trajectory in other layers of the detector. A Combinatorial Kalman Filter is used to perform the final track fit. Hit-based matching criteria are used to identify tracks reconstructed from stau particles. 

When evaluating the optimal hit timing windows and track reconstruction performance, staus are required to be produced with $|\eta| \leq 0.8$, and primarily traverse layers of the tracker barrel. Full optimization of the endcap tracker geometry, timing windows, and track reconstruction is left for future work. Track reconstruction performance, is evaluated using signal events with 10\% of the expected BIB per bunch crossing overlaid. 

To evaluate the physics sensitivity of a search for meta-stable charged particles, two sources of potential background are considered. The first involves tracks reconstructed from a random combination of BIB hits. Events with neutrinos and BIB overlay are used to characterize this background, using a sample of 10 events with 100\% BIB. Standard model muons are considered a second source of potential background. A sample of $10^6$ events with $\mu^+\mu^- \to \mu^+\mu^-$ were generated at leading order using \textsc{MadGraph5\_aMC@NLO} interfaced with \textsc{Pythia8}. 

\section{Hit-timing criteria}
\label{sec:directdet}

Precision timing is an essential handle for rejecting hits from BIB. For each location in the detector, we define the expected time of arrival, $t_0$, for a particle produced at the center of the detector, at the time of the collision $t=0$, and traveling with $\beta=1$. The actual time of arrival for any hit, $t_{\mathrm{hit}}$, can be compared to this expected time of arrival to reject BIB. Standard Model particles produced in collisions will have a corrected hit time $t_{\mathrm{corr}}=t_{\mathrm{hit}}-t_0\sim0$, with smearing due to detector time resolution and a finite beamspot size. The detector resolution, $\sigma_t$, is the dominant contribution to this smearing. The small beamspot size $\sigma_z \sim \SI{1.5}{\milli\meter}$ corresponds to a time spread of ${\sim}\SI{5}{\pico\second}$. Previous studies only consider hits within $-3\sigma_t<t_{\mathrm{corr}}<+5\sigma_t$ for track reconstruction, reducing occupancy by an order of magnitude~\cite{imcc-euro-strat}. We refer to this requirement as the \texttt{Nominal} timing window. 

Slowly moving long-lived particles produce hits that are delayed with respect to the \texttt{Nominal} timing criteria. The accumulated delay is largest for hits in the outermost layers of each sub-detector. To improve hit selection efficiency for slowly moving particles, we define a range of timing criteria shown in Table~\ref{table:time-windows}. \texttt{Loose} criteria are defined to capture $>90\%$ of hits produced by a stau with mass of $4~\mathrm{TeV}$ in the outermost layer of each sub-detector. \texttt{Medium} criteria are defined as a midpoint in occupancy between \texttt{Nominal} and \texttt{Loose} criteria.

\begin{table}[h]
\centering
\begin{tabular}{||c||c|c|c||}
    \hline
    & \multicolumn{3}{c||}{Sub-detector timing window [ns]} \\
    Timing criteria & Vertex Detector & Inner Tracker & Outer Tracker \\
    \hline\hline
     \texttt{Nominal} & -0.09 -- 0.15 & -0.18 -- 0.30 & -0.18 -- 0.30 \\
     \texttt{Medium} & -0.09 -- 0.26 & -0.18 -- 0.50 & -0.18 -- 0.60 \\
     \texttt{Loose} & -0.09 -- 0.40 & -0.18 -- 2.20 & -0.18 -- 10.00 \\
     \hline
\end{tabular}
\caption{The three timing criteria considered by this study, with the corresponding timing window defined for each sub-detector.}
\label{table:time-windows}
\end{table}

\begin{figure}[t]
    \centering
    \includegraphics[width=0.48\linewidth]{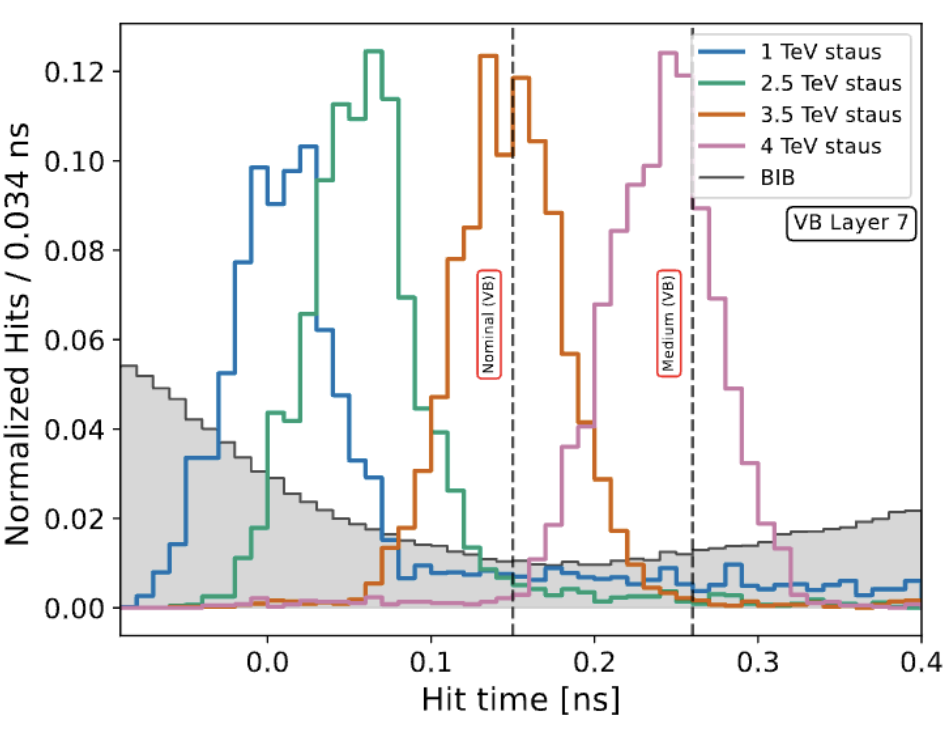}
    \includegraphics[width=0.48\linewidth]{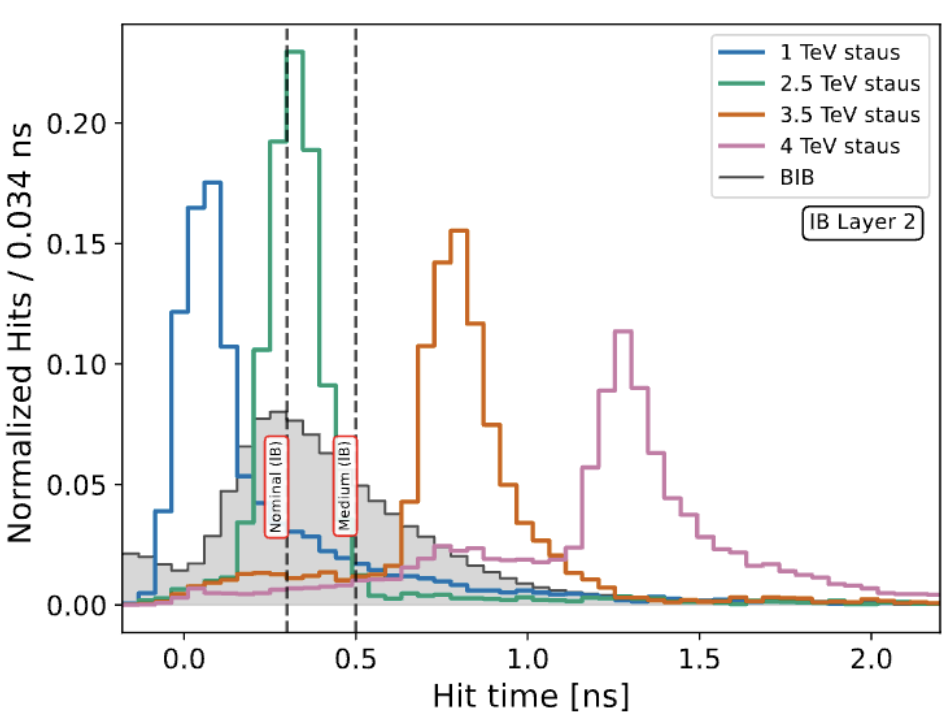}
    \caption{Corrected arrival time distributions for BIB hits and simulated staus with mass $m_{\tilde{\tau}}$ = 1, 2.5, 3.5, and 4 TeV. Vertical dashed lines indicate the boundaries of the \texttt{Nominal} and \texttt{Medium} timing windows. Left: Vertex Barrel Layer 7, Right: Inner Barrel Layer 2.
    }
    \label{fig:arrival_times}
\end{figure}

Figure \ref{fig:arrival_times} shows the corrected hit times for staus with masses of 1, 2.5, 3.5, and 4 TeV in the final layer of the Vertex and Inner Barrels compared to hits from BIB. These distributions are produced before any track reconstruction is performed and are representative of hit-timing profiles in different layers of the detector. Distributions are shown within the \texttt{Loose} timing window for each sub-detector considered. \texttt{Nominal} and \texttt{Medium} timing window cutoffs are displayed by dashed vertical lines. In the Vertex Detector roughly $80\%$ of BIB hits are still excluded by the \texttt{Loose} timing window. In comparison, the bulk of BIB hits occur before $\SI{1}{\nano\second}$ in the Inner and Outer Trackers.

\begin{figure}[b]
  \centering
    \includegraphics[width=0.6\linewidth]{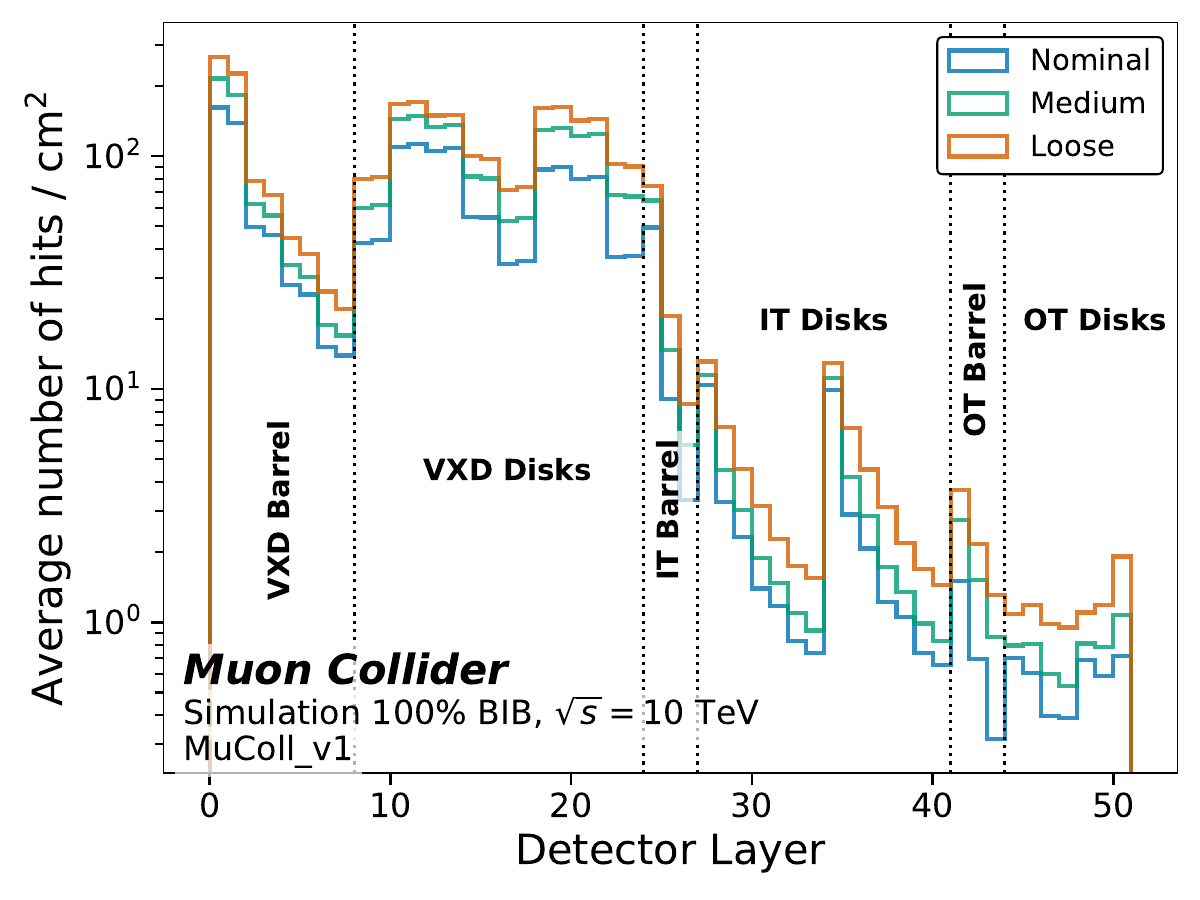}
  \caption{Hits per cm$^2$ in the \textsc{MuColl\_v1} detector with 100$\%$ BIB for each hit-timing window considered.}
  \label{fig:occupancy}
\end{figure}

The occupancy in each layer of the detector due to BIB is shown in Figure~\ref{fig:occupancy} for these three timing windows. Relative to the \texttt{Nominal} case, \texttt{Medium} windows increase occupancy on average by 38$\%$, 35$\%$, and 60$\%$ in the Vertex Detector,  Inner Tracker, and Outer Tracker, respectively. \texttt{Loose} windows increase occupancy by 73$\%$, 105$\%$, and 144$\%$ in the same three sub-detectors. 

\begin{figure}[t]
    \centering
    \includegraphics[width=\linewidth, page=1]{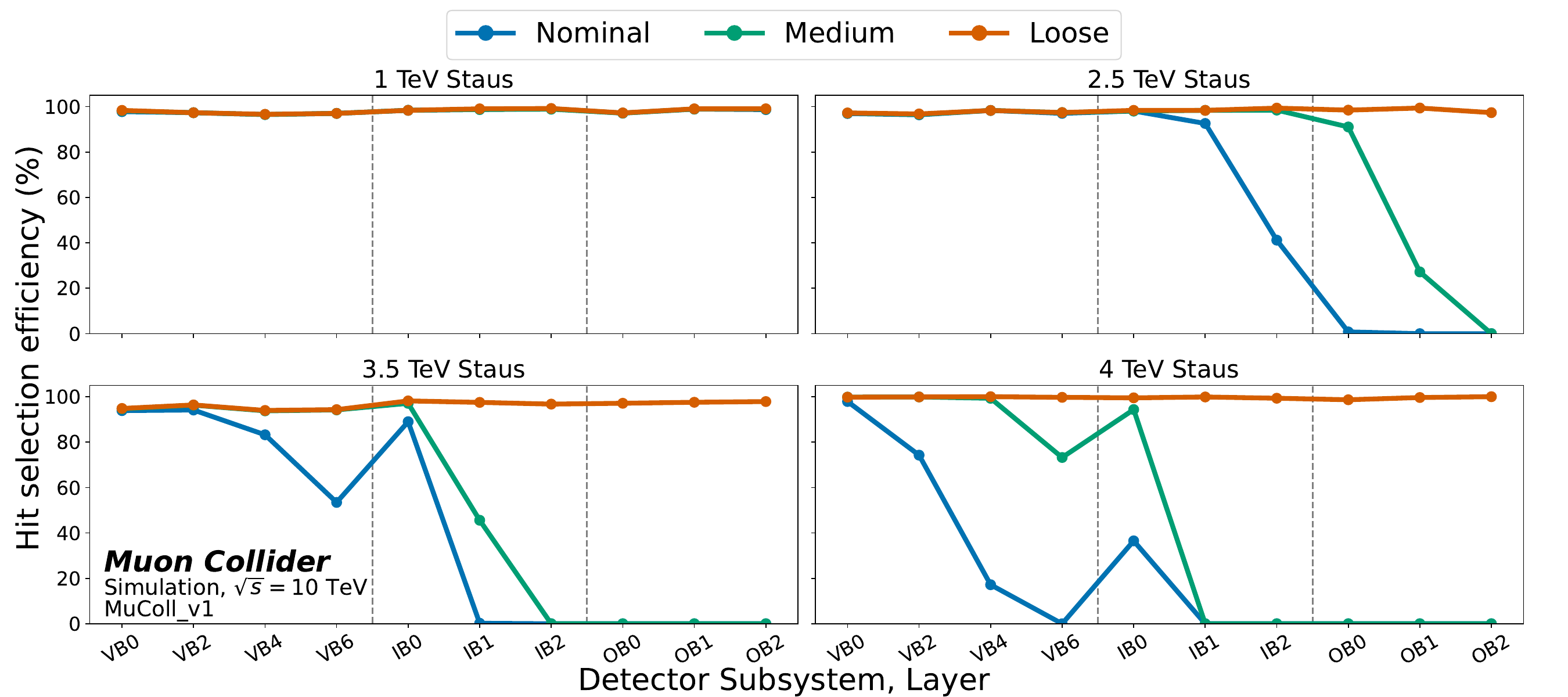}
    \caption{Hit selection efficiency by sub-detector and layer for $m_{\tilde{\tau}}$=1, 2.5, 3.5, and 4 TeV staus independent of whether the hits are matched to tracks.} 
    \label{fig:hit_reco}
\end{figure}

Figure \ref{fig:hit_reco} demonstrates the impact of timing criteria on hit selection efficiency for different stau masses in each layer of the tracker. Hit selection efficiency is defined as  
\(
\epsilon_\text{hit} = N_\text{hits recorded} / N_{\text{hits expected}}
\).
A single hit in a Vertex doublet layer is treated as 0.5 hits, and recorded hits are not required to be associated with a reconstructed track. 

As expected, tighter timing windows result in lower hit selection efficiency. Hit selection efficiency sometimes increases for the first layer of the subsequent sub-detector due to the extended timing window. \texttt{Nominal} timing criteria reject nearly all hits from staus with mass above $1~\mathrm{TeV}$ in the Inner and Outer Tracker. \texttt{Medium} timing windows fully recover hits produced by stau with masses up to $2.5~\mathrm{TeV}$ in the Inner Tracker.   Only the \texttt{Loose} criteria are fully efficient for all masses and all layers considered. 

\FloatBarrier
\section{Track reconstruction performance}
\label{sec:overallreco}

We investigate the impact of loosened hit-timing criteria on track reconstruction. To assess representative background conditions while balancing computational efficiency only $10\%$ of the BIB expected per bunch crossing is simulated. We find that looser timing windows do not noticeably impact the runtime of digitization or track reconstruction. Reconstruction takes on average $\mathcal{O}(\SI{2}{\minute})$ per event with $10\%$ BIB. 

We define the track reconstruction efficiency as
\(
\epsilon_\text{track} = N_{\text{tracks}}/N_{\text{staus}}
\).
Staus are required to traverse all layers of the tracker barrel. All reconstructed tracks are required to be truth-matched to a stau and satisfy $\chi^2/\text{n.d.f}< 3$. No additional criteria are applied to the track transverse momentum with respect to the $p_\mathrm{T}>10~\mathrm{GeV}$ requirement applied to track seeds. 

\begin{table}[h]
\centering
\begin{tabular}{||c||c|c|c|c||}
    \hline
    & \multicolumn{4}{c||}{Track Selection Criteria} \\
    Requirement & $\chi^2 / \mathrm{n.d.f}$ & Vertex Detector & Inner Tracker & Outer Tracker \\
    \hline\hline
     VB &$\leq$ 3 & $\geq$ 3 hits & -- & -- \\
     VB+IB  & $\leq$ 3 & $\geq$ 3 hits & $\geq$ 2 hits & -- \\
     VB+IB+OB & $\leq$ 3 & $\geq$ 3 hits & $\geq$ 2 hits & $\geq$ 2 hits\\
     \hline
\end{tabular}
\caption{Track selection criteria}
\label{table:track-crit}
\end{table}

We investigate the impact of requiring hits in different sub-detectors by defining three track selection criteria,  summarized in Table~\ref{table:track-crit}. The loosest possible selection only requires hits in the Vertex Barrel.  Stricter criteria require hits in subsequent sub-detectors, resulting in a lower rate of fake tracks as well as improved momentum and velocity resolution. 

\begin{figure}[t]
    \centering
    \includegraphics[width=\linewidth, page=1]{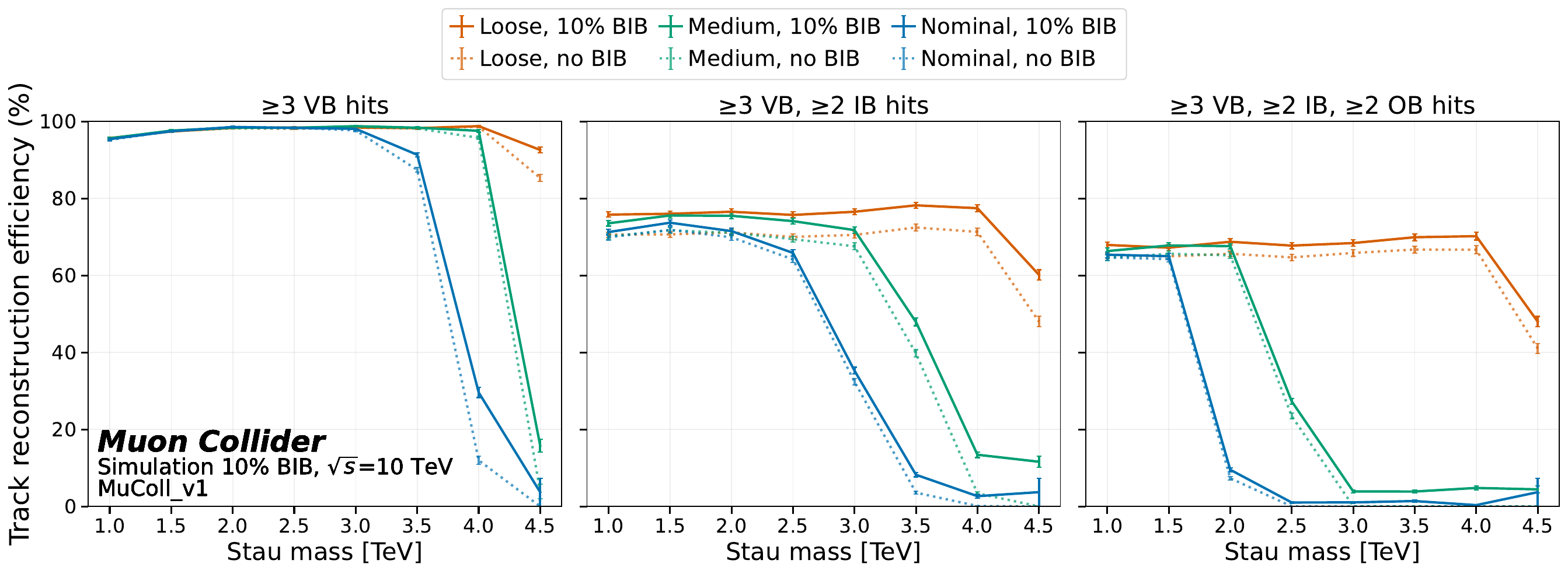} 
    \caption{Track reconstruction efficiency for staus with and without BIB overlay. Efficiencies are shown for all time windows and each track selection criteria.} 
    \label{fig:full_eff}
\end{figure}

Figure \ref{fig:full_eff} shows the reconstruction efficiency for tracks satisfying these criteria for different stau masses and timing windows, with and without BIB overlay. Tracking efficiency suffers for heavier masses and tighter timing windows, as hits from slowly-moving particles will arrive too late to fall within timing acceptance windows. This effect is worsened in the Inner and Outer Tracker, where track reconstruction is nearly impossible for $m_\text{stau} > 2.5$ TeV with the \texttt{Nominal} and \texttt{Medium} timing windows. The \texttt{Loose} windows allow for $\sim$70$\%$ track reconstruction efficiency when requiring tracks to have hits in the Outer Tracker. Track reconstruction efficiency is inflated by a few percent in the presence of BIB. 

The \texttt{Loose} timing window also improves track purity despite the increased BIB occupancy by retaining additional signal hits, maintaining an average purity above 95\% for stau masses up to 4 TeV. The resulting longer tracks and larger lever arm also improve the momentum and velocity resolution, with velocity resolution reaching ${\sim}1\%$ when hits are required in all tracker sub-detectors. The presence of BIB degrades both momentum and velocity resolution by introducing incorrectly associated hits, particularly for higher mass staus reconstructed with tighter timing windows (Appendix~\ref{sec:appTrackReco}). 






\FloatBarrier
\section{Search for meta-stable charged particles}
\label{sec:search}

We investigate how modified hit-timing windows and track reconstruction translate to sensitivity to meta-stable charged particles. Potential sources of background include tracks reconstructed from random combinations of BIB hits and Standard Model muons.  

Samples with 100\% BIB were used to optimize selections and estimate the background contribution from tracks reconstructed from random combinations of BIB hits. BIB events were digitized with each of the three timing windows. 
Standard Model muons are expected to be produced with high-$p_{\mathrm{T}}$ and travel at nearly the speed of light. Their reconstruction should not be affected by timing window acceptance. Therefore, muons were only digitized using the \texttt{Nominal} window for simplicity.

\begin{figure}[htbp]
    \centering

    \begin{subfigure}{0.49\textwidth}
        \centering
        \includegraphics[page=2, width=\linewidth]{plots/sig_bib_mu_leadsub.pdf}
        \caption{Reconstructed \pT{}}
        \label{fig:bkg-truth-pt}
    \end{subfigure}
    \hfill
    \begin{subfigure}{0.49\textwidth}
        \centering
        \includegraphics[page=3, width=\linewidth]{plots/sig_bib_mu_leadsub.pdf}
        \caption{Reconstructed \(\beta\)}
        \label{fig:bkg-truth-beta}
    \end{subfigure}

    \vspace{0.5em}

    \begin{subfigure}{0.49\textwidth}
        \centering
        \includegraphics[page=1, width=\linewidth]{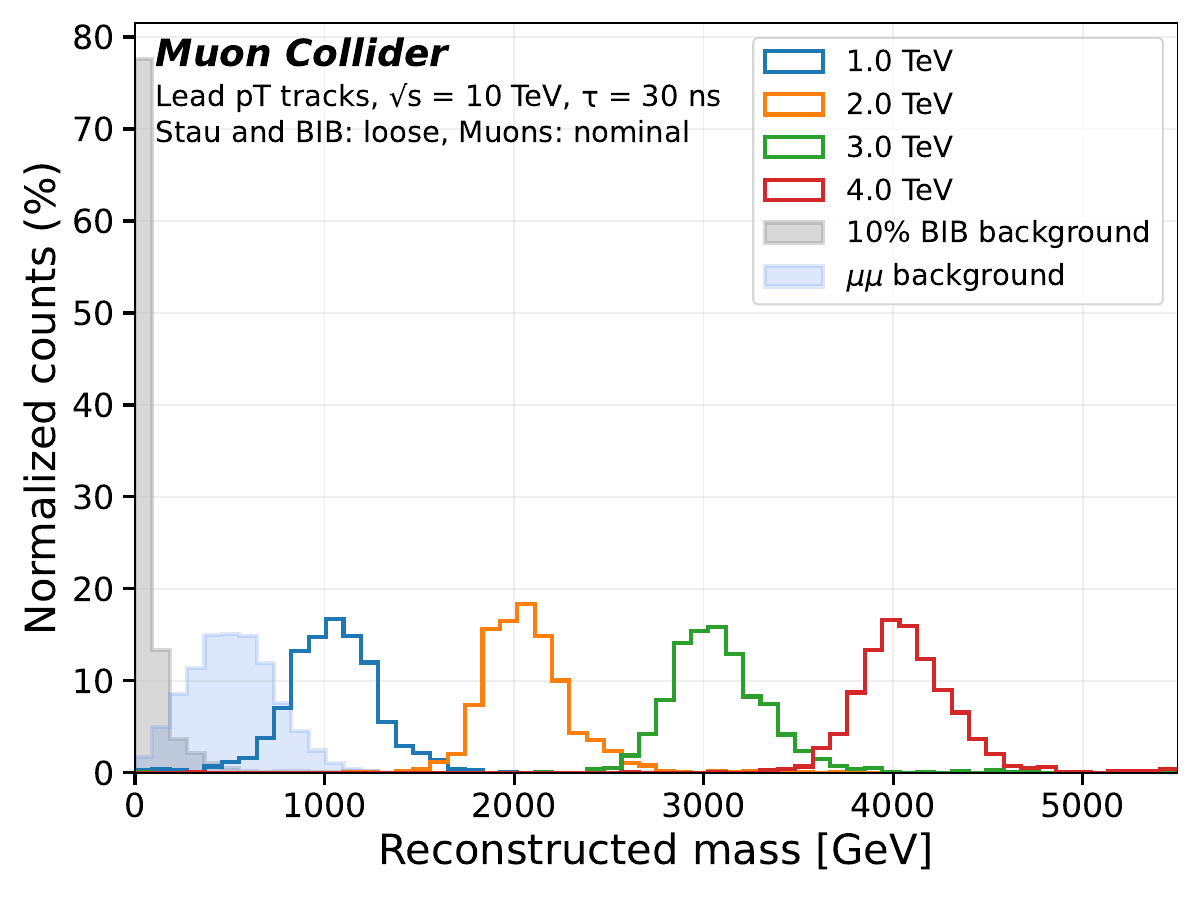}
        \caption{Reconstructed mass}
        \label{fig:bkg-truth-mass}
    \end{subfigure}

    \caption{Reconstructed track \pT{}, \(\beta\), and mass for the leading \pT{} track in \(\mu^+\mu^-\) background, BIB, and selected stau signals after quality $|\eta|\leq0.8$, $\chi^{2}/\mathrm{n.d.f.}<3$, and $\geq$ 3 VB, 2 IB, 2 OB hit requirements. BIB distributions are shown using simulations with \(10\%\) BIB overlay in the \texttt{Loose} window and are used to illustrate the shapes of reconstructed BIB-track observables.}
    \label{fig:bkg-truth}
\end{figure}

Reconstructed properties of tracks for BIB, Standard Model muons, and a subset of stau signals are overlaid in Figure \ref{fig:bkg-truth}. To enable sufficient statistics, a sample with 10\% BIB overlay is used to demonstrate the shape of the BIB observables. BIB tracks reconstructed with a \texttt{Loose} timing window typically have low $p_\mathrm{T}$, a wide distribution of $\beta$ values, and a low reconstructed mass ($m_{\mathrm{reco}}$). Muon tracks are typically reconstructed with $\beta \approx 1$, high \pT{}, and $m_{\mathrm{reco}} \approx 0$. Due to a small tail of muons reconstructed with $\beta < 1$ values, a Gaussian-like peak around $m_{\mathrm{reco}} \approx 500$ GeV is observed. Reconstructed stau mass, $p_{\mathrm{T}}$, and $\beta$ vary as expected for pair-produced particles, but generally have higher $m_{\mathrm{reco}}$ and $p_{\mathrm{T}}$ with lower $\beta$ than backgrounds. 

Various selections are applied to reduce BIB and Standard Model muons to negligible levels while maximizing signal acceptance. Track-level requirements are first applied to ensure that each reconstructed track is well measured and consistent with a single particle trajectory. Event-level selections are then imposed on the leading and subleading tracks to distinguish the stau signal from the remaining backgrounds.

\subsection{Track selection}

Signal and background tracks are required to lie within the central tracker acceptance, $|\eta|\leq0.8$, and satisfy the VB+IB+OB track selection criteria outlined in Table \ref{table:track-crit}. Tracks must also fall within $\pT{}<10~\mathrm{TeV}$ to reject unphysical momentum measurements. To ensure hit timing consistency, the weighted root mean square of the hit-timing residuals relative to the fitted track velocity must satisfy $w_{\mathrm{RMS}}<1.4$.

\begin{figure}[h]
    \centering
    \includegraphics[width=0.8\linewidth]{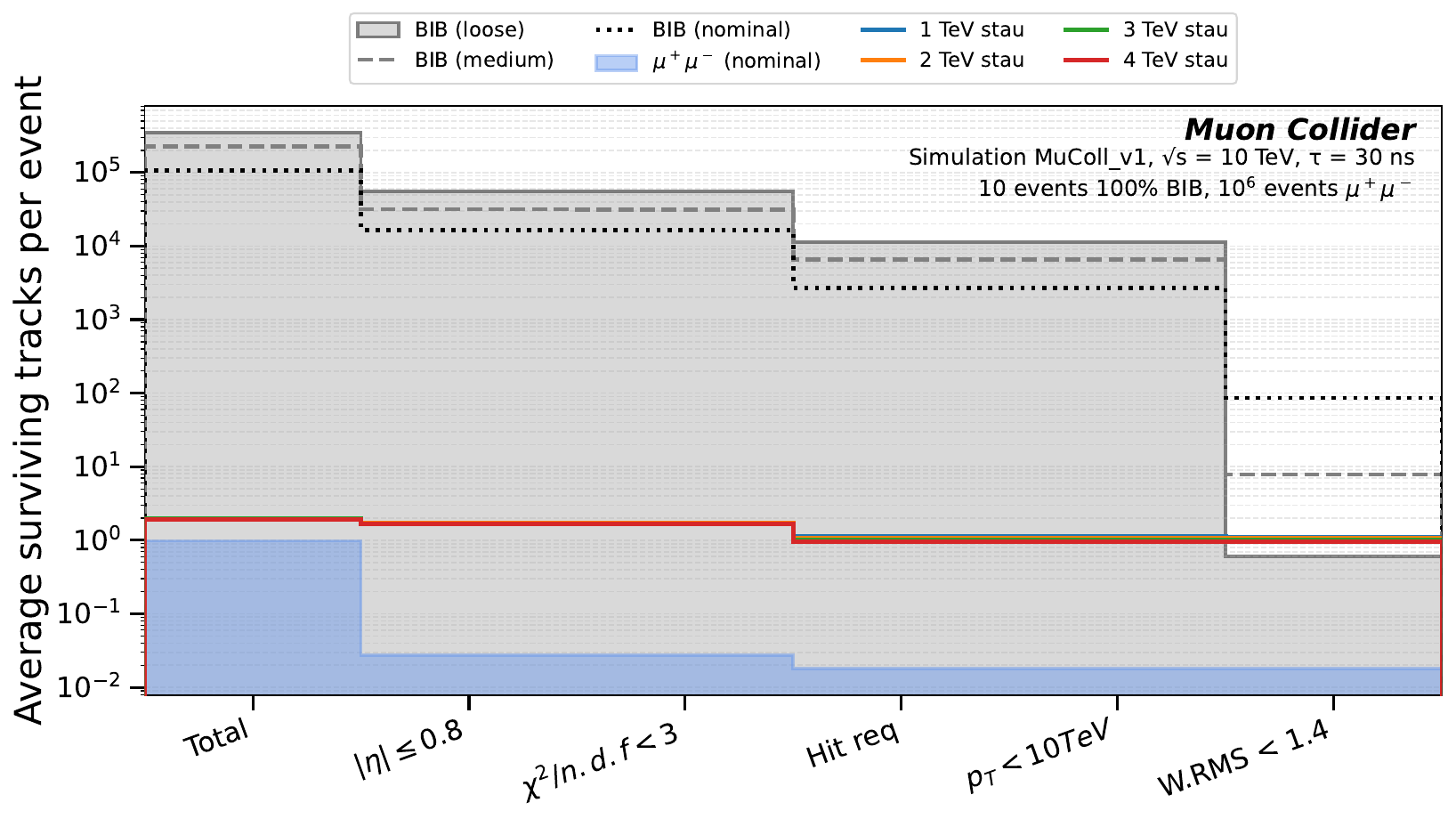}
    \caption{Cumulative track-level cutflow visualization for muon, BIB, and \texttt{Loose} window stau tracks. Cuts are applied from left to right, with each bin representing the average number of tracks per event remaining after the selection denoted on the $x$ axis, for 10 BIB, $10^6$ muon, and 2500 signal events per stau mass hypothesis.}
    \label{fig:cutflow-vis}
\end{figure}

For BIB, the track-level selections strongly suppress the number of surviving tracks in all timing windows, as seen in Figure \ref{fig:cutflow-vis}. The geometric acceptance and hit requirements provide the largest reductions before the timing-consistency cut is applied. It is worth noting that more tracks survive the final \(w_{\mathrm{RMS}}<1.4\) requirement in the tighter timing windows, despite the fact that the tighter windows contain fewer total reconstructed BIB tracks. This behavior is expected from the definition of \(w_{\mathrm{RMS}}\). 
\texttt{Nominal} and \texttt{Medium} timing windows restrict allowed hit times to a narrower interval. Surviving tracks have small timing residuals and therefore more easily pass the \(w_{\mathrm{RMS}}\) requirement.

\subsection{Event selection}

From the remaining tracks in every event, only the leading and subleading $\pT{}$ tracks are kept. These two tracks are then subject to a series of stau-mass dependent cuts, as described in Table \ref{table:mass-cuts}. For each stau mass hypothesis, both the leading and subleading tracks are independently tested against the corresponding set of mass-dependent selection criteria, effectively treating each hypothesis as a separate analysis.

\begin{table}[h]
\centering
\begin{tabular}{||c||c|c|c||}
    \hline
    & \multicolumn{3}{c||}{Selection Criteria} \\
    $m_{\tilde{\tau}}$ [TeV] & $p_\text{T}^{\min}$ [GeV] & $\beta$ window & $m_{\mathrm{reco}}$ window [GeV]\\
    \hline\hline
    1.0 & 1500 & 0.95--0.983 & 500--2000 \\
    1.5 & 2000 & 0.90--0.98  & 1000--2000 \\
    2.0 & 2000 & 0.85--0.95  & 1500--3000 \\
    2.5 & 2000 & 0.80--0.90  & 2000--4000 \\
    3.0 & 2000 & 0.70--0.85  & 2000--6000 \\
    3.5 & 2000 & 0.65--0.80  & 2900--5000 \\
    4.0 & 1900 & 0.55--0.70  & 3200--5500 \\
    4.5 & 1400 & 0.40--0.55  & 3500--6000 \\
    \hline
\end{tabular}
\caption{Mass-dependent kinematic selection criteria applied to both leading and subleading tracks for signal, BIB, and muon backgrounds, as  optimized for 30 ns stau efficiency.}
\label{table:mass-cuts}
\end{table}

\begin{figure}
    \centering
    \includegraphics[width=0.8\linewidth]{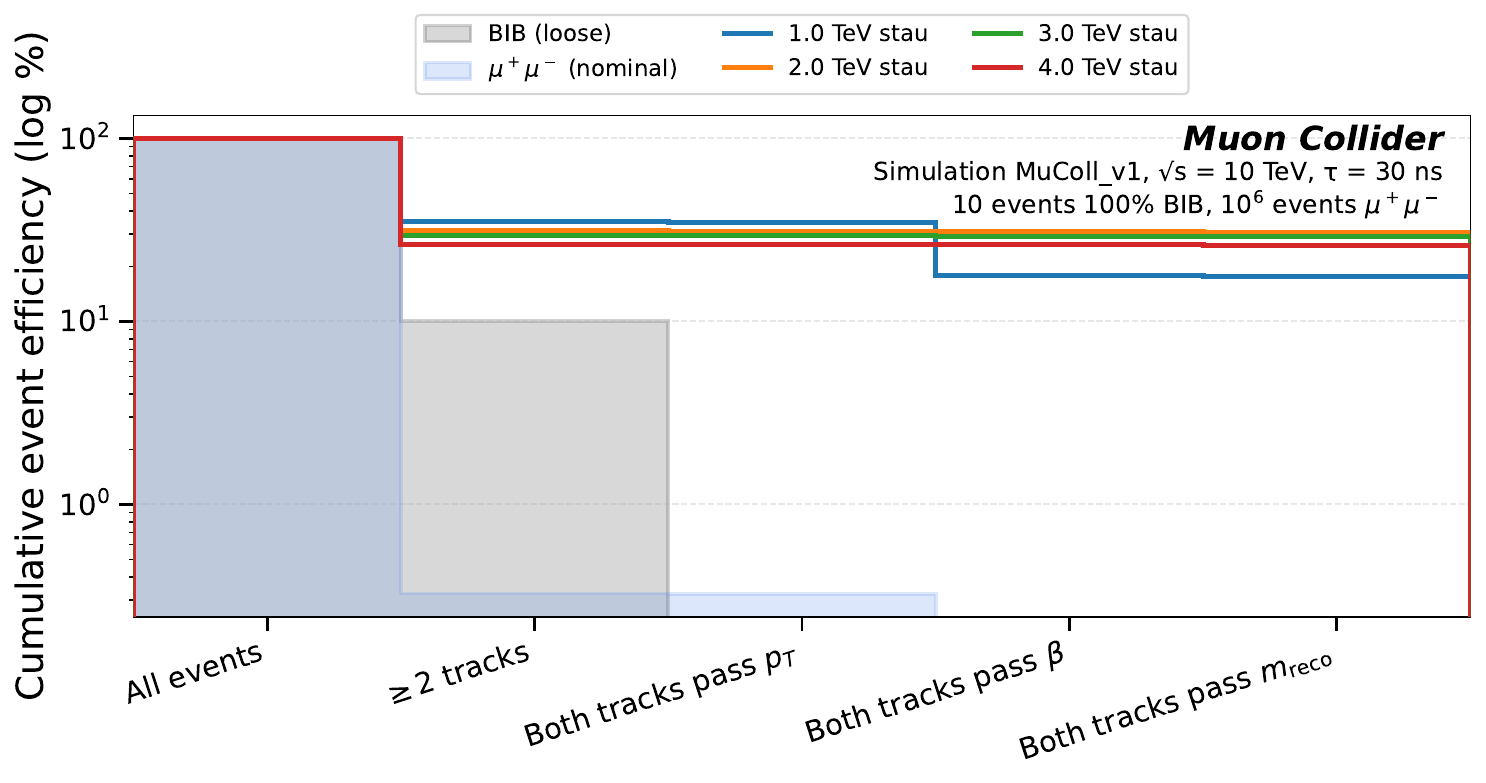}
    \caption{Cumulative event-level cutflow for representative 30 ns stau samples and the BIB and muon backgrounds. Signal and BIB samples use the \texttt{Loose} window, while the muon sample uses the \texttt{Nominal} window. The background cutflows are evaluated using the 3 TeV mass-dependent selection.}
    \label{fig:event-level-cutflow}
\end{figure}

The event-level cutflow is partially shown in Figure~\ref{fig:event-level-cutflow}. Only a representative subset of all signal and background data is shown, as the observed qualitative trends are consistent across all mass and lifetime configurations. The requirement of two reconstructed tracks after applying the track-level requirements is the most stringent selection, leading to a substantial reduction in event yield in both signal and background. This is especially true for more massive staus, as decreased tracking efficiency takes effect. The transverse momentum (\pT{}) requirement further suppresses background contributions, eliminating most of BIB and a large fraction of muon pairs while retaining a substantial fraction of the signal. The $\beta$ selection removes the remaining background events in the simulated samples, though it also removes a fraction of the $1~\mathrm{TeV}$ stau signal. A reconstructed-mass ($m_{\mathrm{reco}}$) selection window is then applied as an additional safeguard against residual background, with only a small impact on the stau signal efficiency.

\subsection{Background Estimation}

After applying the full sequence of track-level
requirements and the subsequent event-level selection, no Standard Model muon events survived the stacked cutflow. Because expected yields are normalized to the target integrated luminosity of $10~\mathrm{ab}^{-1}=10^7~\mathrm{pb}^{-1}$ expected for five years of data-taking at a muon collider \cite{imcc-euro-strat}, the absence of surviving events in this high-statistics sample indicates that the residual muon background is negligible within the statistical precision of the simulation. We therefore treat the muon background as negligible for the purposes of the following sensitivity estimate.

Unlike the muon background, the BIB background is limited to 10 fully simulated events. We perform a per-track ABCD estimate to obtain stronger statistical evidence for BIB rejection~\cite{ATL-PHYS-PUB-2026-014}.
Using the larger number of reconstructed BIB tracks available at the track-level, the ABCD method estimates the probability for an individual BIB track to pass the signal requirements. This per-track probability is then propagated under a Poisson model to the probability that an event contains at least two passing tracks, in a sample much larger than the 10 fully simulated BIB events.

The two selections used for the per-track ABCD method include the timing consistency requirement, \[
X \equiv \left(w_{\mathrm{RMS}} < 1.4\right),
\] and the combined stau mass-dependent selection criteria denoted in Table \ref{table:mass-cuts}.
\[
Y \equiv
\left(\pT{} > p_\text{T}^{\min}\right)
\cap
\left(\beta^{\min}<\beta<\beta^{\max}\right)
\cap
\left(m_{\mathrm{reco}}^{\min}
<m_{\mathrm{reco}}
<m_{\mathrm{reco}}^{\max}\right).
\]

This ABCD method is applied after a baseline track quality selection, defined by $|\eta|\leq0.8$, $\chi^2/\mathrm{n.d.f.}<3$, and hit requirements of at least 3 Vertex Barrel hits, 2 Inner Barrel hits, and 2 Outer Barrel hits. Applying the ABCD procedure after these baseline quality requirements ensures that the estimate reflects the discriminating power of the final timing and kinematic selections, X and Y, which constitute the dominant background rejection criteria in the signal selection.

The ABCD estimate first predicts the number of BIB tracks in the signal-like region A as \[
A_{\mathrm{pred}} = \frac{BC}{D}.
\] Under the assumption that fake tracks are rare and approximately independently distributed across events, the number of passing tracks per event is modeled by a Poisson distribution with mean, \[
\lambda = \frac{A_{\mathrm{pred}}}{N_{\mathrm{events}}},
\] where $N_{\mathrm{events}}=10$ is the number of simulated 100\% BIB events per timing window. The probability that an event contains at least two passing tracks is calculated as \[
P_{\geq 2} = 1-e^{-\lambda}(1+\lambda),
\] and the expected number of such events normalized to the expected $1.5 \times 10^{12}$ events at a 10 TeV muon collider is \[
N_{\mathrm{fake}{\geq2}} = (1.5 \times
10^{12})P_{\geq2}.\]

\begin{figure}[h!]
    \centering
    \includegraphics[width=0.75\linewidth]{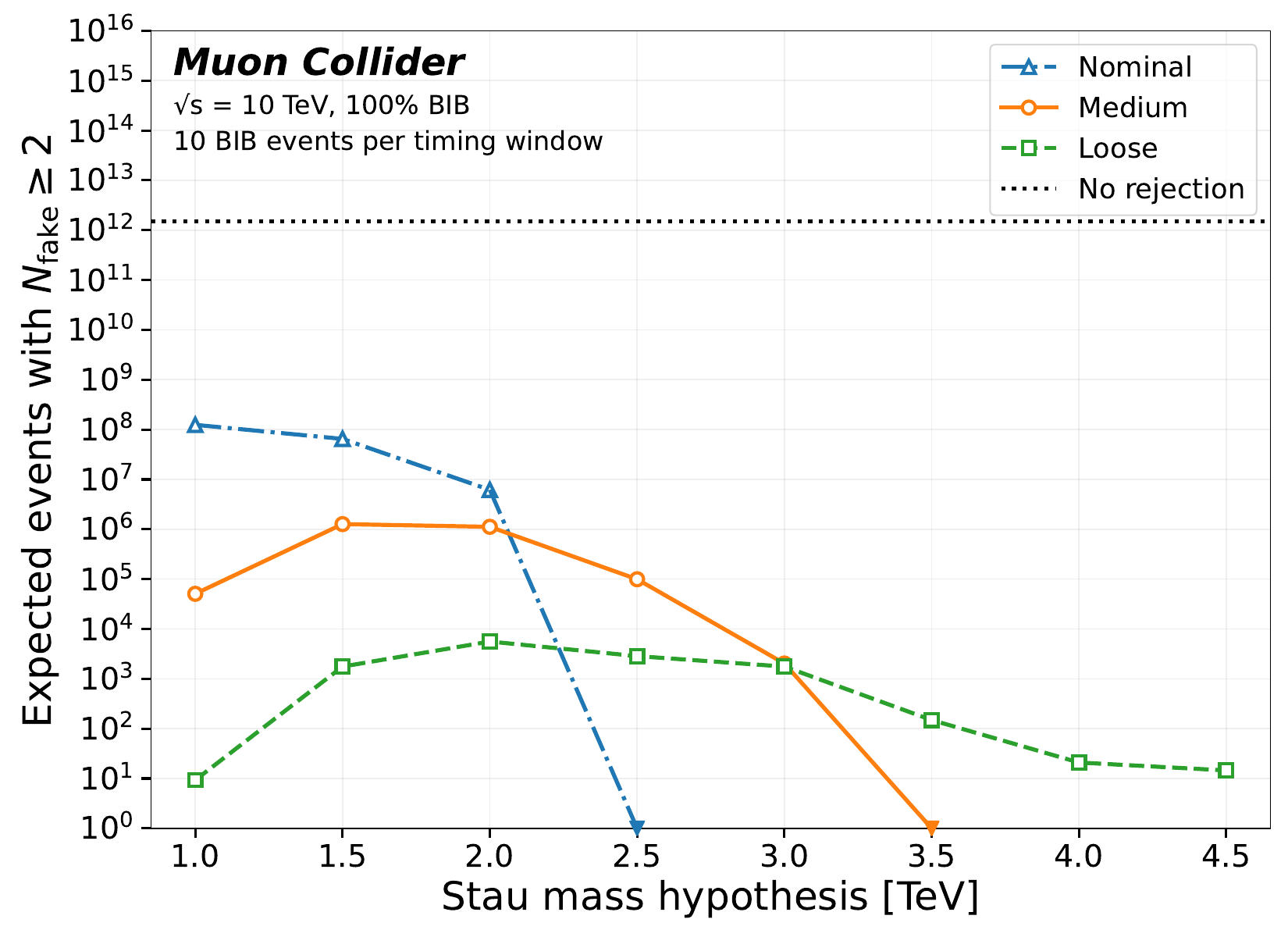}
    \caption{Expected number of BIB events containing at least two candidate tracks satisfying baseline quality and stau mass-dependent  selections, shown for the \texttt{Loose}, \texttt{Medium}, and \texttt{Nominal}
timing windows. Event yields are obtained from the ABCD predictions assuming independently distributed fake tracks and normalized to $1.5\times10^{12}$ collision events. The dotted line indicates the BIB event yield with no background rejection, and points displayed at the lower plotting boundary correspond to predictions of zero events.}
    \label{fig:abcd_estimate}
\end{figure}

The resulting estimates of the number of events containing at least two fake BIB tracks, $N_{\mathrm{fake}\geq2}$, are shown in Figure \ref{fig:abcd_estimate} for the three timing windows as a function of the stau mass hypothesis. For all timing windows and mass hypotheses considered, the underlying ABCD estimate predicts far fewer than one BIB track in the signal-like region per 10 simulated 100\% BIB events, demonstrating strong track-level background suppression. 
The \texttt{Loose} timing window produces the smallest predicted background yield for the 1 TeV stau mass hypothesis, while its largest yields occur for intermediate masses near 2--3 TeV. This behavior follows from the overlap between the mass-dependent $\beta$ selection windows and the reconstructed BIB $\beta$ distribution, whose characteristic shape is illustrated using the 10\% \texttt{Loose} BIB sample in Figure \ref{fig:bkg-truth-beta}. For the \texttt{Medium} and \texttt{Nominal} timing windows, the number of predicted fake BIB events decreases to zero at higher mass hypotheses because their narrower, more relativistic BIB $\beta$ distributions do not extend into the lower $\beta$ signal regions in Table \ref{table:mass-cuts}. However, as shown in Figure \ref{fig:full_eff}, these tighter timing windows also have negligible reconstruction efficiency for the corresponding higher mass stau signals.



\begin{figure}[h!]
    \centering
    \includegraphics[width=0.75\linewidth]{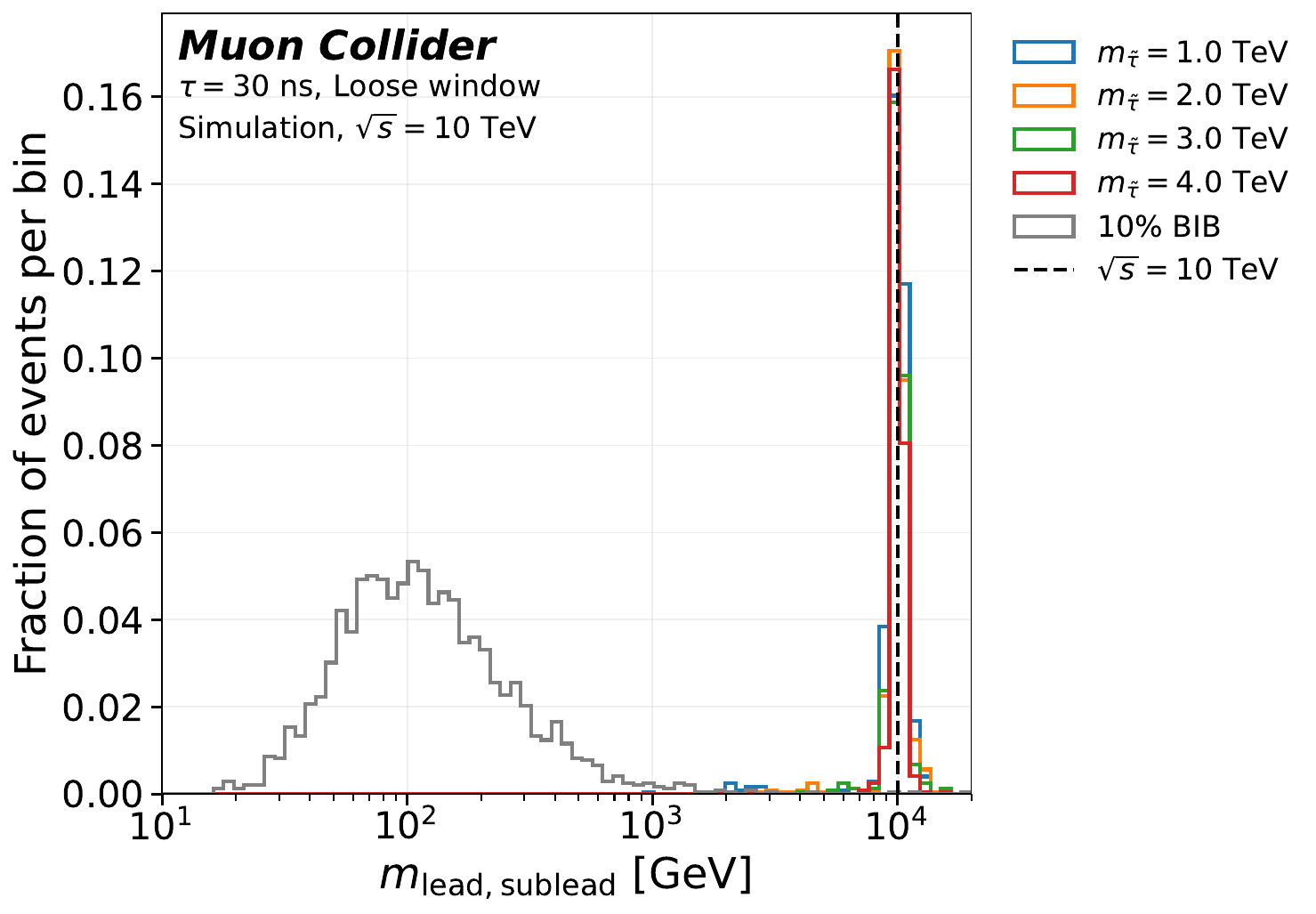}
    \caption{Ditrack invariant-mass distributions of the leading and subleading \pT{} candidate tracks for selected 30 ns stau signal samples and 10\% BIB using the \texttt{Loose} timing window. The stau signal is concentrated near $\sqrt{s}=10$ TeV, while reconstructed BIB events populate the lower mass region}
    \label{fig:ditrack_mass}
\end{figure}

Additional event-level observables can provide further discrimination beyond the selections utilized in the ABCD estimate. Figure \ref{fig:ditrack_mass} depicts the invariant mass of the leading and subleading candidate tracks with clear discrimination between stau signal and BIB. Stau pair production is concentrated near the center-of-mass energy 10 TeV while reconstructed BIB tracks are concentrated at substantially lower invariant masses. Due to the per-track nature of the ABCD method, the rejection provided by a ditrack invariant mass requirement cannot be incorporated reliably into the ABCD background estimation, but the strong shape separation observed suggests that a ditrack invariant mass selection could provide substantial additional BIB rejection.

Moreover, further BIB suppression is expected from ongoing and anticipated improvements to track reconstruction. In particular, stricter rejection of outlier hits, track extrapolation between sub-detectors, and 4D timing aware reconstruction are expected to substantially reduce the formation of fake tracks. 

The results of the ABCD estimate, in combination with the discriminating power of the di-track invariant mass and expected improvements to track reconstruction, provide evidence that the BIB contribution can be suppressed to negligible levels. We therefore evaluate sensitivity to the stau signal under a zero--background assumption.

\subsection{Results}

Under this target zero--background scenario, the projected sensitivity is evaluated for an integrated luminosity of \(\mathcal{L}=10~\mathrm{ab}^{-1}\). Signal cross sections, \(\sigma_{\text{theory}}\), for a \(\sqrt{s}=10~\mathrm{TeV}\)  are displayed in Figure~\ref{fig:cross-sections}.

A 95\% confidence-level exclusion is approximated by requiring an expected signal yield of \(s > 3\) events, corresponding to the Poisson threshold for excluding a signal at 95\% CL in the zero-background limit. The corresponding expected cross-section sensitivity is obtained from
\[
\sigma_{95} = \frac{s}{\epsilon \cdot \mathcal{L}},
\]
with \(s=3\). If no simulated signal events pass the full selection, the corresponding efficiency is zero and no finite cross-section limit is quoted.

\begin{figure}[h]
    \centering
    \includegraphics[width=0.8\linewidth]{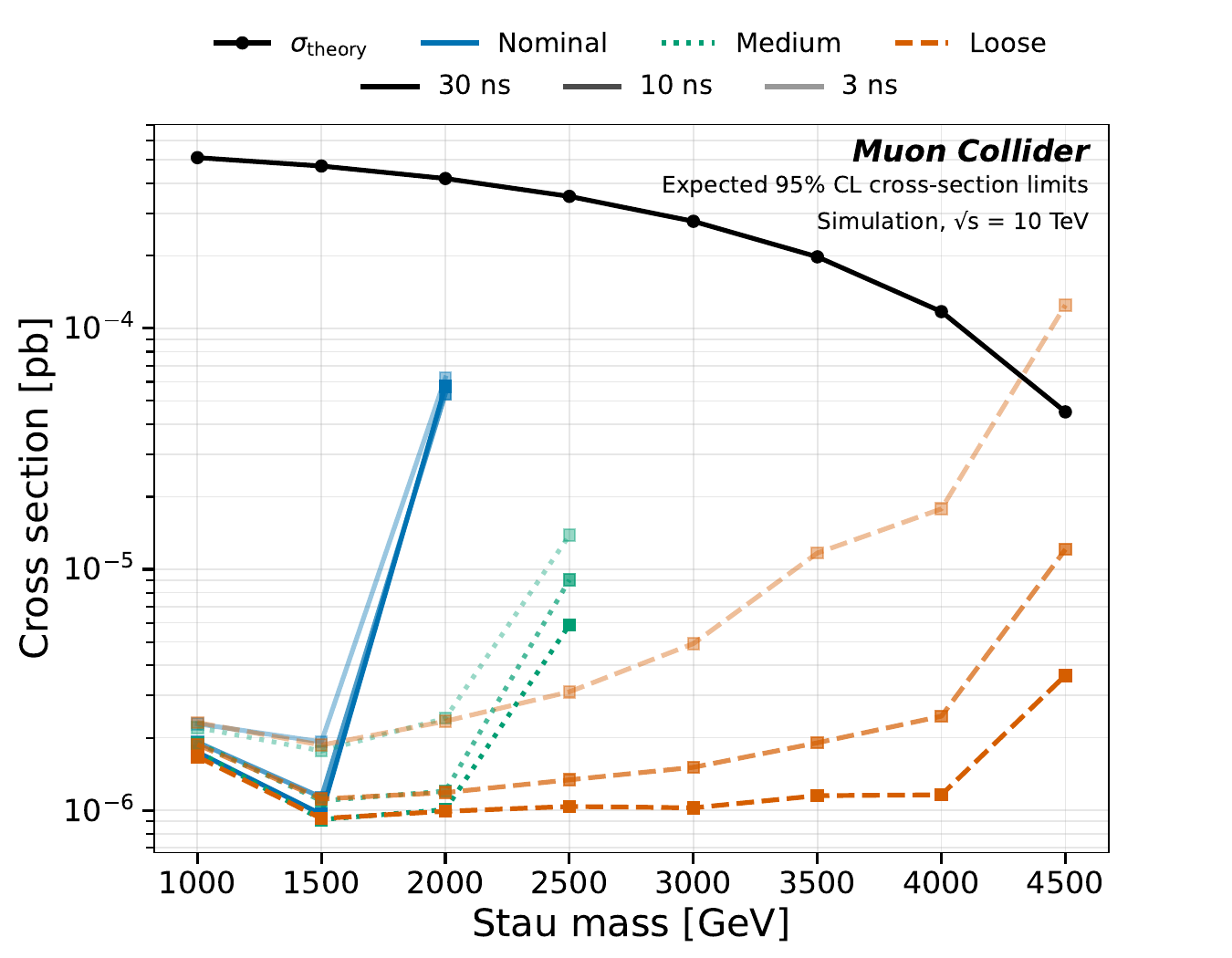}
    \caption{Expected 95\% confidence-level cross-section limits for directly simulated stau lifetimes. The black curve shows the MadGraph theory prediction.}
    \label{fig:xs-lim}
\end{figure}

The resulting cross-section limits are shown in Figure~\ref{fig:xs-lim} for the directly simulated lifetimes of 3, 10, and 30~ns, with the MadGraph theory prediction overlaid for comparison. The \texttt{Nominal} and \texttt{Medium} timing windows lose sensitivity at high stau mass as the reconstruction efficiency falls to zero, while the \texttt{Loose} timing window maintains sensitivity to heavier and longer-lived staus by preserving delayed hits from slowly moving particles.

\begin{figure}[h]
    \centering
    \includegraphics[width=0.8\linewidth]{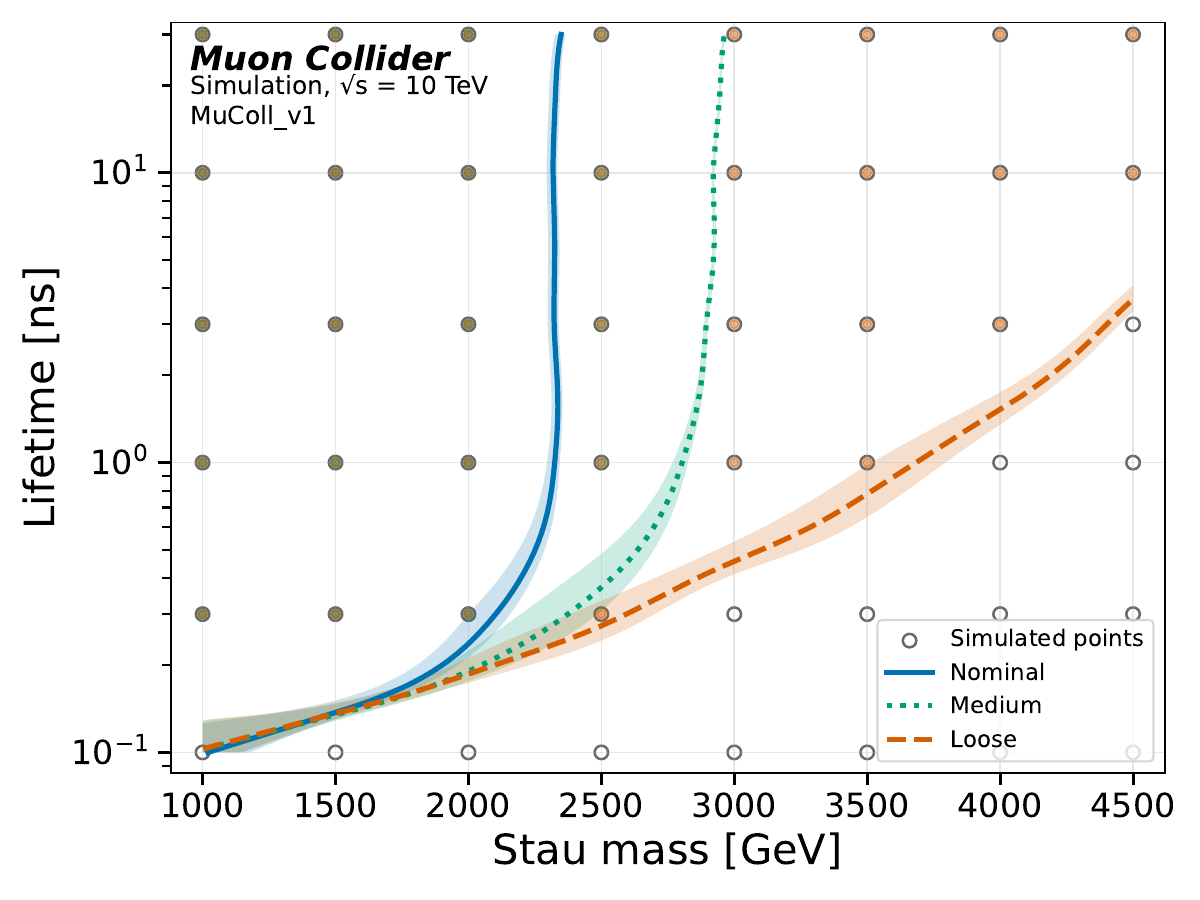}
    \caption{Projected 95\% confidence-level exclusion contours in the stau mass--lifetime plane. Contours are obtained by interpolating the expected signal yield and locating the \(s=3\) boundary. Shaded bands show the effect of shifting the simulated signal yields by their \(\pm1\sigma\) statistical uncertainties before repeating the interpolation. Open markers indicate the simulated grid points used as inputs to the interpolation.}
    \label{fig:exclusion-contour}
\end{figure} 

The resulting projected exclusion contours are shown in Figure~\ref{fig:exclusion-contour}. The \texttt{Loose} timing window provides the strongest projected reach, excluding a substantially larger region of parameter space than the \texttt{Medium} or \texttt{Nominal} timing windows. These projections demonstrate that the detector timing configuration has a direct impact on the physics reach for slowly moving long-lived particles. The \texttt{Nominal} timing window provides strong BIB rejection but removes delayed signal hits from heavy staus, leading to sharply reduced sensitivity. The \texttt{Medium} timing window partially recovers this acceptance, while the \texttt{Loose} timing window gives the best overall performance across the studied mass--lifetime plane.

\FloatBarrier
\section{Conclusions}

The presence of beam-induced-background at a muon collider poses significant challenges for detector design, reconstruction, and physics performance. While narrow hit-timing requirements effectively reduce BIB occupancy, they also reject delayed hits from slowly moving charged long-lived particles. Using long-lived staus as a benchmark, we demonstrate that extending tracker timing acceptance is necessary to preserve sensitivity to heavy LLPs at a $10~\mathrm{TeV}$ muon collider while maintaining reasonable occupancy.

Of the three timing configurations considered, the \texttt{Loose} window recovers delayed hits across the tracker without producing a prohibitive increase in occupancy, and yields track reconstruction efficiencies of approximately 70\% for staus traversing the full barrel. The additional accepted hits also improve track purity and momentum and velocity resolution, particularly for high-mass staus.

Several improvements to track reconstruction are needed to fully exploit the tracker’s physics potential. Improved extrapolation between subsystems would preserve track continuity and improve momentum resolution, while a more capable refit algorithm could remove BIB hits incompatible with the reconstructed track hypothesis. Alternative approaches should also be explored: seeding in the low-occupancy outer layers could reduce the number of seeds, accelerate reconstruction, and enable displaced-track reconstruction. Timing-aware algorithms that extract a particle’s $\beta$ could further suppress outer-layer occupancy while retaining delayed charged long-lived-particle signals~\cite{casarsa-4dtracking}. Alongside improved reconstruction algorithms, future sensor-level BIB hit rejections are being explored, including doublet layers and smart pixels~\cite{parpillon2024smartpixelsinpixelai}.


We also demonstrate that track-level quality and timing requirements, followed by mass-dependent selections on the transverse momentum and velocity of the two leading tracks, strongly suppress both BIB and high \pT{} muon backgrounds. In the \texttt{Loose} window, no simulated background events survive the full selection, and a per-track ABCD estimate demonstrates large suppression of BIB. These results motivate evaluating the projected sensitivity under a zero-background approximation, while larger BIB samples will be needed to validate this assumption more precisely. 

Under the zero-background assumption, the cross section limits and mass--lifetime exclusion contours demonstrate that hit timing windows have a direct impact on physics reach. The \texttt{Nominal} and \texttt{Medium} windows lose sensitivity at high stau mass, while the \texttt{Loose} window retains sensitivity over all proposed stau masses with a minimum lifetime around 3 ns. 

Muon collider experiment design and expected BIB will evolve substantially in the coming years. Multiple 10 TeV detector concepts are under development ~\cite{maia-concept, music}, and improved tracker geometries are under study. The overall level of BIB expected in the detector will also evolve as the
collider ring and accelerator lattice design continue to develop. As these designs mature, it will be critical to revisit and extend the present study. Ensuring next-generation collider experiments are capable of capturing long-lived particle signatures is essential for maximizing discovery reach. 

\acknowledgments

We are grateful to the International Muon Collider Collaboration, whose software development was crucial to enable this work. We thank S. Pagan Griso, F. Meloni, A. Badea, I. Hirsch, and N. Virani for useful discussions and suggestions. 

This work utilizes resources from the University of Chicago and the Enrico Fermi Institute. K.F. Di Petrillo is supported by the National Science Foundation (NSF) CAREER program through Award \#2443370 and the Neubauer Family Assistant Professor Program. B. Rosser and M. Larson are supported by NSF Award \#2310094. K.F. Di Petrillo and B. Rosser are additionally supported by the Simons Foundation. K. Wandrisco is a UChicago Quad Undergraduate Research Scholars Program.

This research was conducted using services provided by the Open Science Grid, which is supported by NSF awards \#2030508 and \#1836650. We are particularly grateful to Pascal Paschos for his support.


\bibliographystyle{JHEP}
\bibliography{biblio.bib}

\FloatBarrier
\appendix
\newpage

\section{Track reconstruction performance}
\label{sec:appTrackReco}

\begin{figure}[t]
    \centering
    \includegraphics[width=0.75\linewidth, page=2]{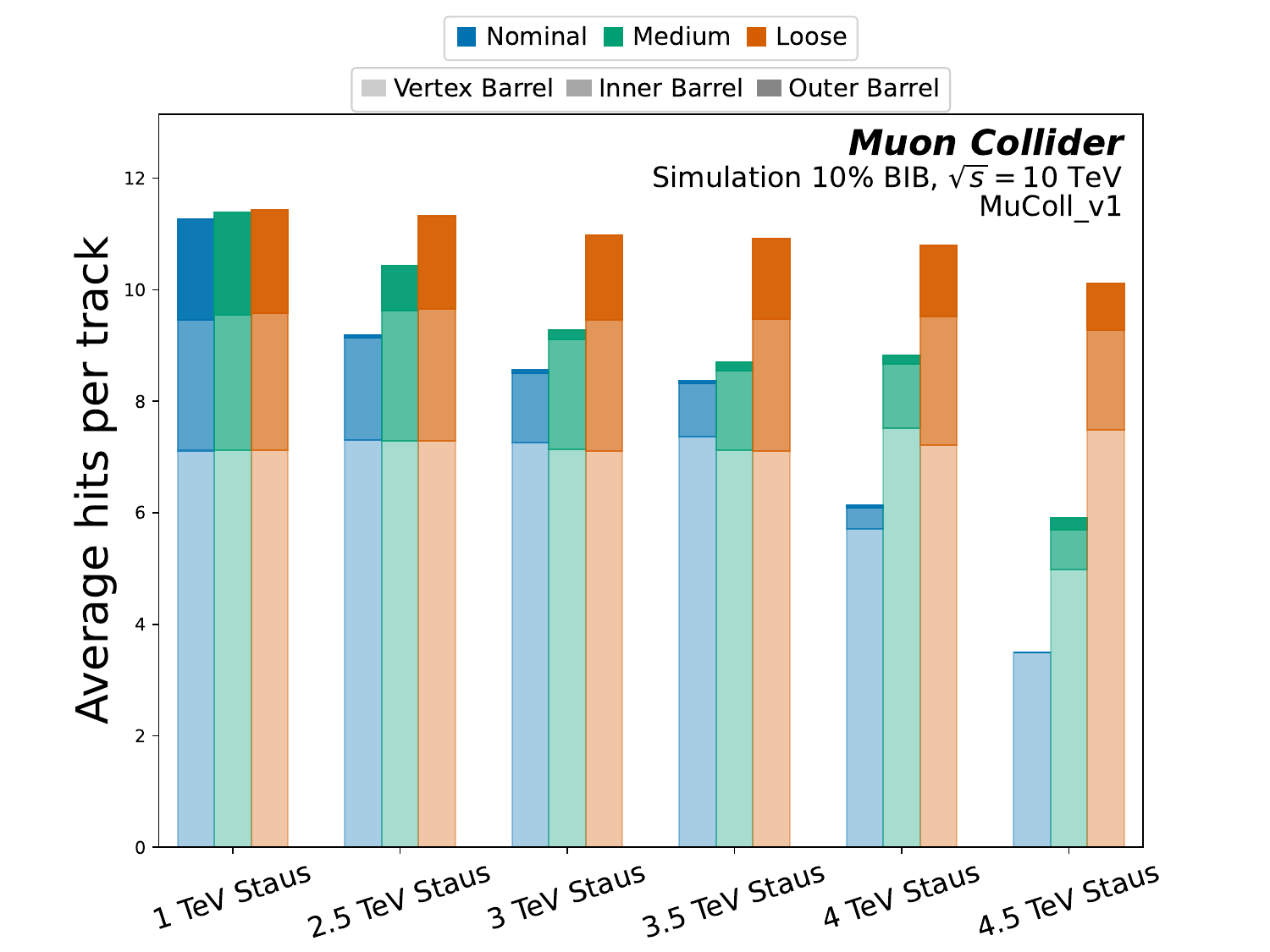}
    \caption{Average number of hits per track in each sub-detector, with 10$\%$ BIB overlay.}
    \label{fig:hits_per_track}
\end{figure}

Figure~\ref{fig:hits_per_track} shows the average number of hits in each sub-detector for each stau mass and timing window considered. For massive staus and tighter timing windows, tracks are shorter on average and consist primarily of Vertex Detector hits.

\begin{figure}[b]
    \centering
    \includegraphics[width=\linewidth, page=1]{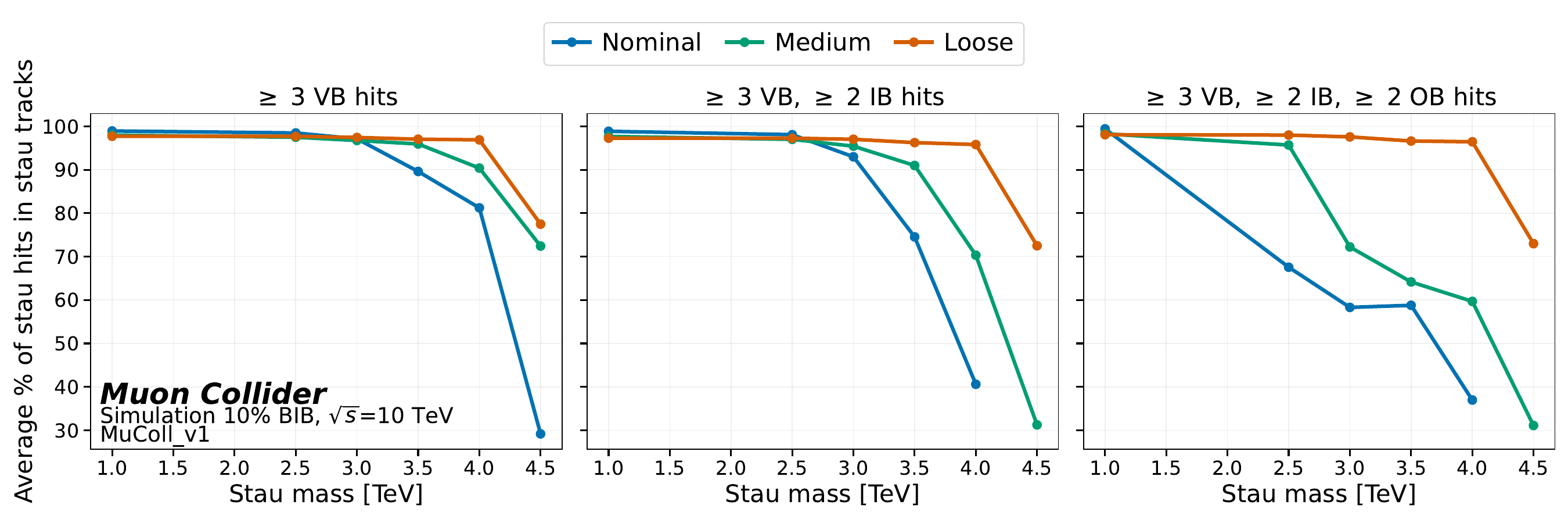}
    \caption{Stau track purity for each mass, window, and track acceptance requirement.}
    \label{fig:percent_stau}
\end{figure}

Track purity, or the fraction of track hits which are truth-matched to a stau, is shown in Figure~\ref{fig:percent_stau}, for the three different track selection criteria and timing windows. Purity is lower for tighter timing windows and higher mass staus, reaching as low as $30\%$ for the \texttt{Nominal} window. Investigation into individual tracks reveals that when signal hits fail timing criteria, the track reconstruction algorithm can attach BIB hits instead. The \texttt{Loose} timing window greatly improves track purity for staus with masses up to $4~\mathrm{TeV}$, with an average purity above $95\%$.

\begin{figure}[h]
    \centering
    \includegraphics[width=\linewidth, page=3]{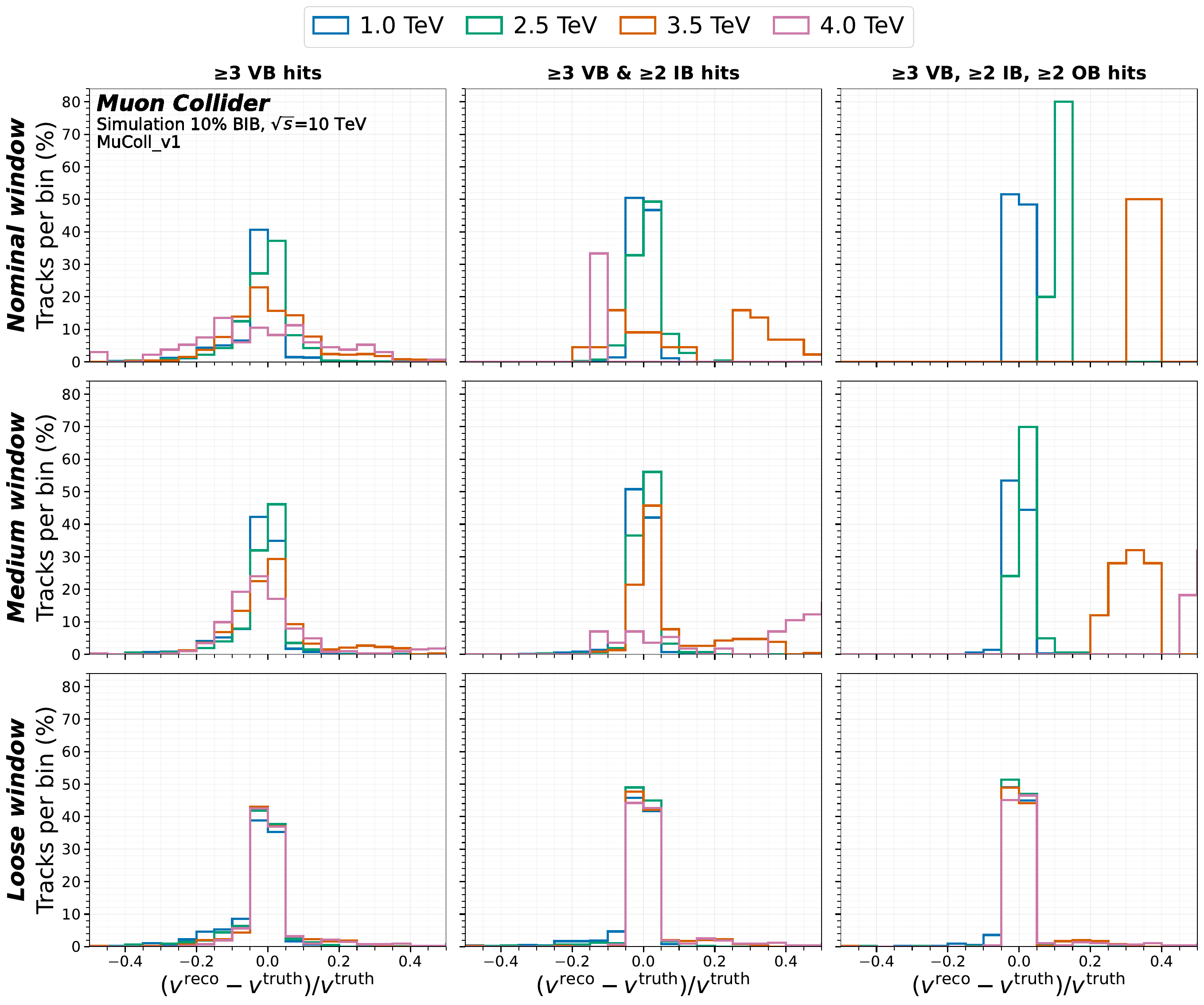}
    \caption{Relative velocity residual for each hit-timing window and track selection criteria for $m_{\text{stau}}$=1, 2.5, 3.5, 4 TeV with $10\%$ BIB.}
    \label{fig:velo_res}
\end{figure}

A particle's velocity is determined by performing a linear fit to the time and space coordinates of each hit in a track. Relative velocity residuals are shown in Figure~\ref{fig:velo_res} for events reconstructed with BIB. The velocity resolution is highly dependent on track purity. For the \texttt{Nominal} timing window, the resolution degrades for higher mass staus and longer track lengths. In the \texttt{Loose} timing window, higher track purity corresponds to better velocity resolution. When hits are required in all sub-detectors, the velocity resolution is  ${\sim}1\%$ for all stau masses.  
 
\begin{figure}[t]
    \centering
    \includegraphics[width=\linewidth, page=2]{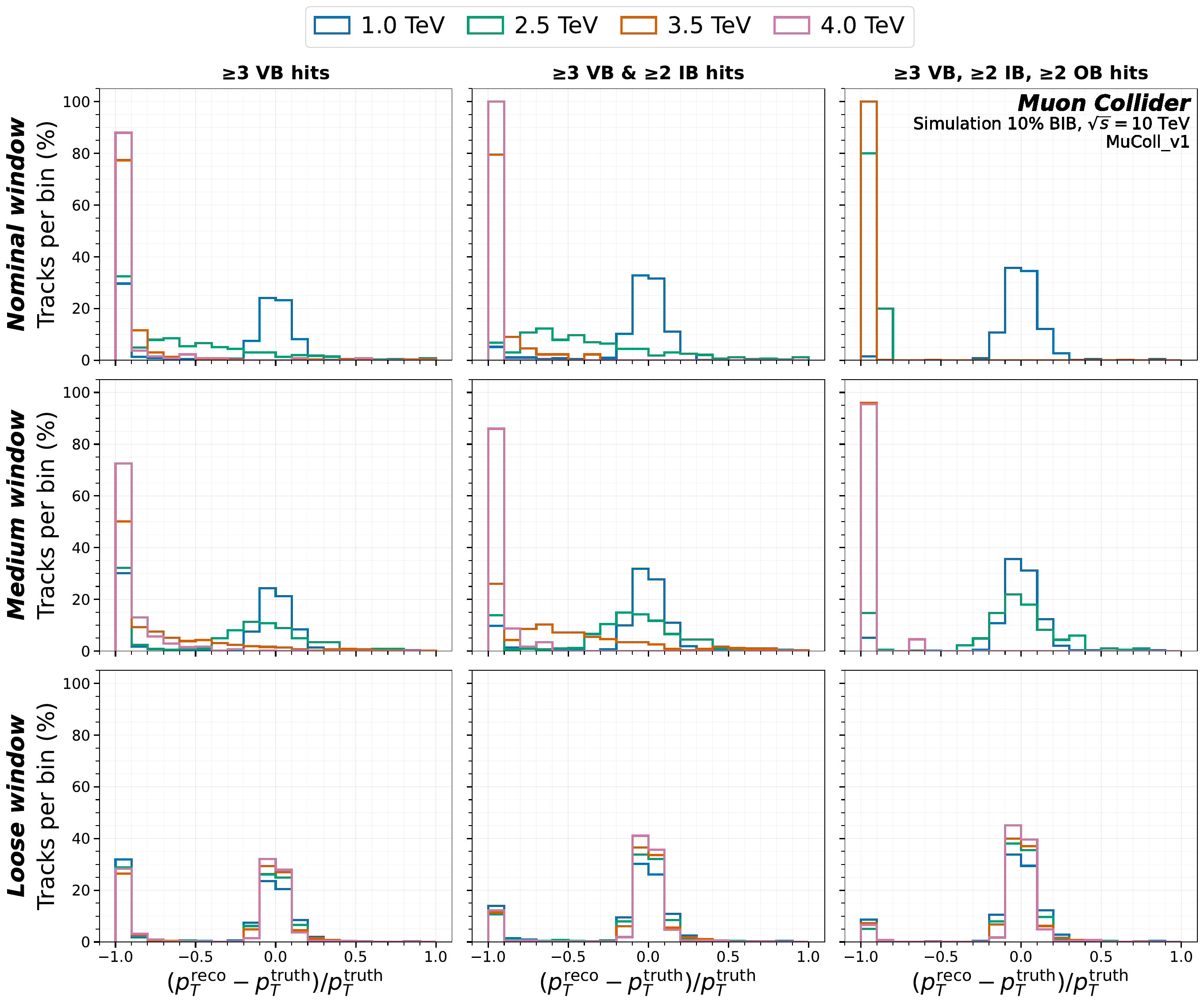}
    \caption{Relative $p_{\mathrm{T}}$ residual for each hit-timing window and track selection criteria, for $m_{\text{stau}}$=1, 2.5, 3.5, 4 TeV with $10\%$ BIB.}
    \label{fig:pT_res}
\end{figure}

We also investigate momentum resolution in the presence of BIB. Relative momentum residuals are shown in Figure~\ref{fig:pT_res}. We find that the reconstructed \pT{} is often underestimated, resulting in a peak around $(p_\mathrm{T}^{\text{reco}} - p_\mathrm{T}^{\text{truth}})/p_\mathrm{T}^{\text{truth}} \approx-1$. This effect is most prominent for higher mass staus reconstructed with tighter timing windows, which have the lowest track purity. The addition of hits to an otherwise straight, high $p_\mathrm{T}$ trajectory bias the curvature measurement and result in lower reconstructed momentum. Momentum resolution improves with the \texttt{Loose} timing window due to the larger lever arm, additional hits on track, and higher track purity.

Even within the \texttt{Loose} window and most stringent track requirements, the reconstructed $p_{\mathrm{T}}$ is sometimes under-estimated. This effect is also observed in samples without BIB overlay. We find that ACTS often classifies hits in the outer tracker layers as outliers during track extrapolation. It should also be noted that the definition of track $\chi^2$ does not include contribution from such outliers. Additional studies are provided in Appendix~\ref{sec:nobib}.

\section{Track reconstruction performance without BIB}
\label{sec:nobib}

\begin{figure}[h]
    \centering
    \includegraphics[width=0.8\linewidth, page=2]{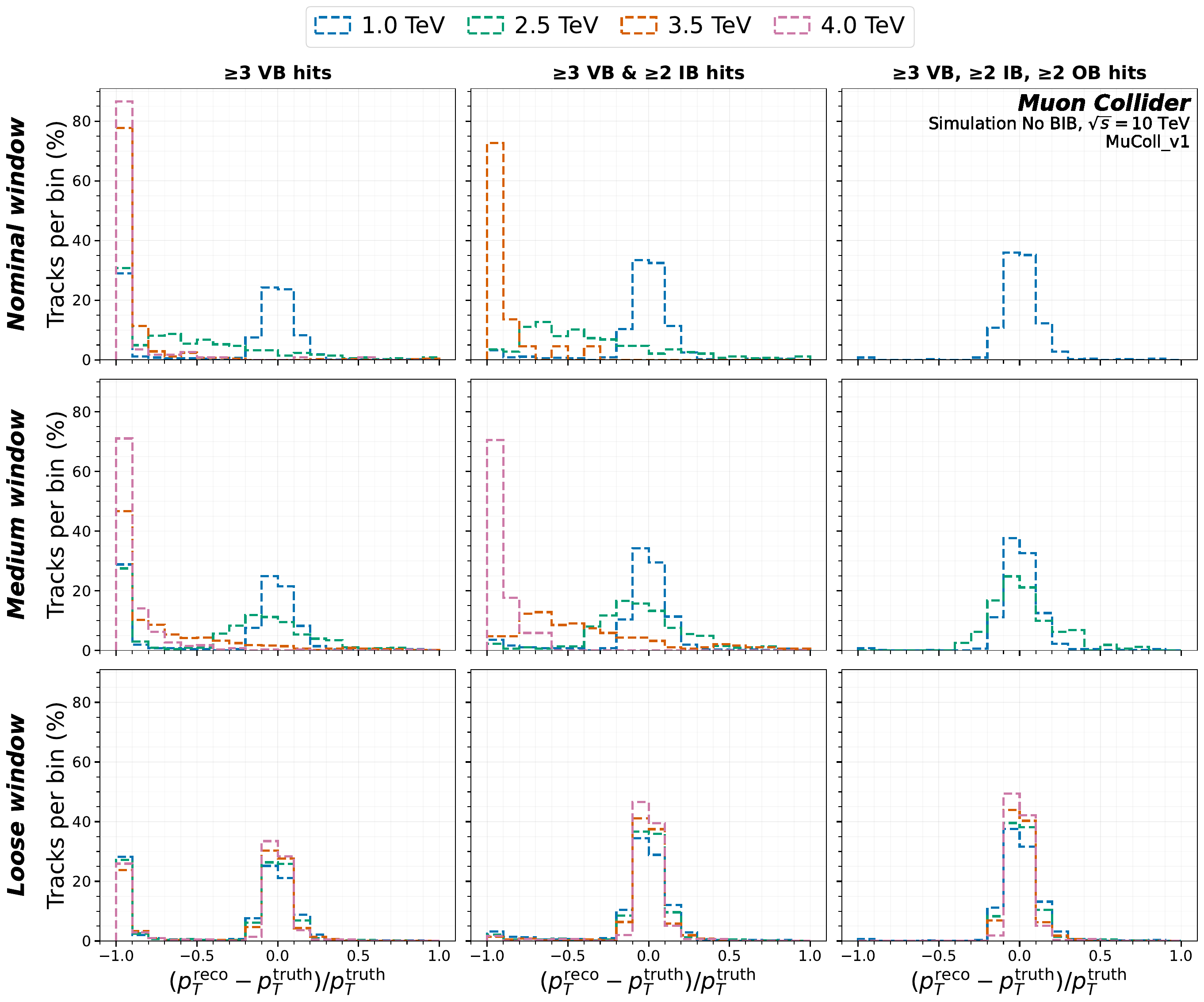}
    \caption{Relative $p_\mathrm{T}$ residual for tracks reconstructed without BIB overlay.}
    \label{fig:pt-res_nobib}
\end{figure}

\begin{figure}[h]
    \centering
    \includegraphics[width=0.8\linewidth, page=3]{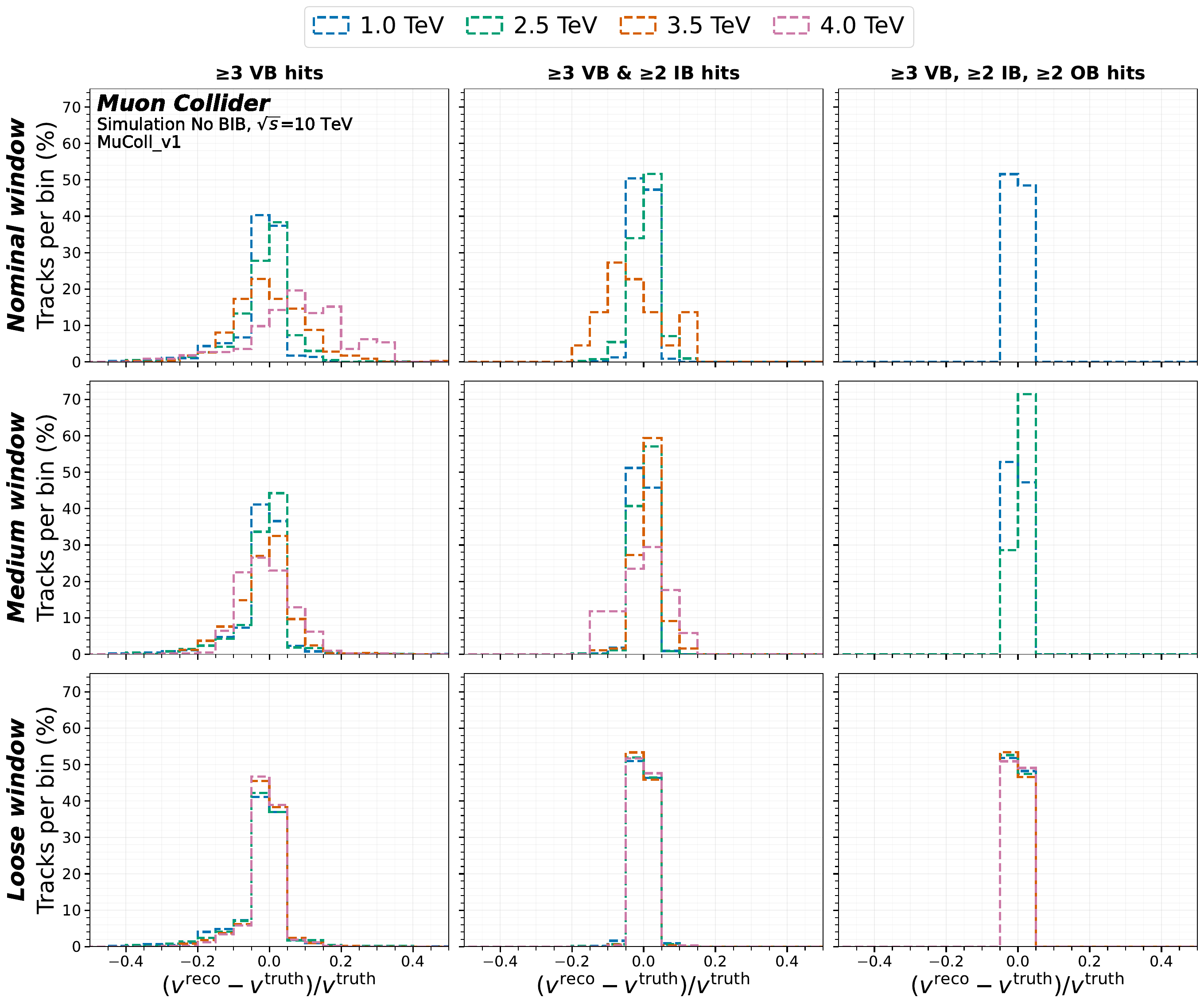}
    \caption{Relative velocity residual for tracks reconstructed without BIB overlay.}
    \label{fig:velo-res_nobib}
\end{figure}

For tracks produced without BIB overlay, relative $p_\mathrm{T}$ and velocity residuals are displayed in Figure~\ref{fig:pt-res_nobib} and Figure~\ref{fig:velo-res_nobib}, respectively. Track reconstruction without BIB decreases the likelihood of severely under-estimating a particle's $p_\mathrm{T}$ and overestimating its velocity.

\end{document}